\pdfoutput=1

\documentclass{acmsiggraph}
\usepackage[margin=1.9in]{geometry} 
\usepackage{dsfont}
\usepackage{wrapfig}
\usepackage{upgreek}
\usepackage{overpic}
\usepackage{amsmath,amsfonts,amssymb,amsthm}

\usepackage{ifpdf}
\DeclareGraphicsExtensions{.png, .jpeg}

\usepackage{color}
\definecolor{lightgray}{rgb}{0.64, 0.64, 0.64}

\pdfauthor{none}
\keywords{none}
\title{Modeling and Correspondence of Topologically Complex 3D Shapes\footnote{Technical Report, School of Computing Science, Simon Fraser University, SFU-CMPT TR 2015-55-2}}
\author{Ibraheem Alhashim}

\begin{document}
	\maketitle
	\begin{abstract}

3D shape creation and modeling remains a challenging task especially for novice users. 
Many methods in the field of computer graphics have been proposed to automate the often 
repetitive and precise operations needed during the modeling of detailed shapes. This 
report surveys different approaches of shape modeling and correspondence especially for shapes exhibiting topological complexity. We focus on methods designed to help generate or process shapes with 
large number of interconnected components often found in man-made shapes. 
We first discuss a variety of modeling techniques, that leverage existing shapes, in easy 
to use creative modeling systems. We then discuss possible correspondence strategies for 
topologically different shapes as it is a requirement for such systems. 
Finally, we look at different shape representations and tools that facilitate 
the modification of shape topology and we focus on those particularly useful in free-form 3D modeling.

\end{abstract}

	
	\section{Introduction} \label{chap:intro}

The creation of quality 3D shapes have been traditionally exclusive to expert designers and artists in different fields ranging from industrial design to visual effects. Recent advancement in the usability of modeling software, in conjunction with powerful computing hardware, have opened up computer aided design and rapid prototyping for both professional and novice users. Experts often use primitive shape tools, such as line or pen tools, to define 2D outlines or use solid primitives and polygon modification tools for free-form 3D modeling. The more advanced tools that can simplify the modeling process include customized scripts that execute replication or deformation based on some predefined rule set. For a novice user who wants to prototype ideas in her mind, it is much easier to modify on existing shapes than to start from scratch. Furthermore, content creators often draw inspiration from many different partial components of existing examples. The mixing and blending of available parts help in rapid generation of creative designs.

Recently available software facilities creative design by utilizing existing large datasets of tailored content or hand made templates. Such systems have been well established in the fields of audio production and in publication and web design. However, in the visual-content field it has yet to expand beyond proposed research projects or specialized 3D character creation software \cite{Poser2014,DAZ2014}. With the enormous availability of visual media as afforded by the Internet, users are now able to draw inspiration form very large datasets of visual content. Recent proposed image composition systems have simplified the conversion of freehand sketches and annotations into photomontages extracted from online images \cite{Szeliski2010}. Similar ideas have also been applied in manual sketch-based 3D part assembly systems that relay on large datasets \cite{Xie2013}. Such data-driven tools that enable creative design have yet to be widely adapted or commercialized.

\begin{figure}[t!]
	\centering
	\includegraphics[width=0.99\linewidth]{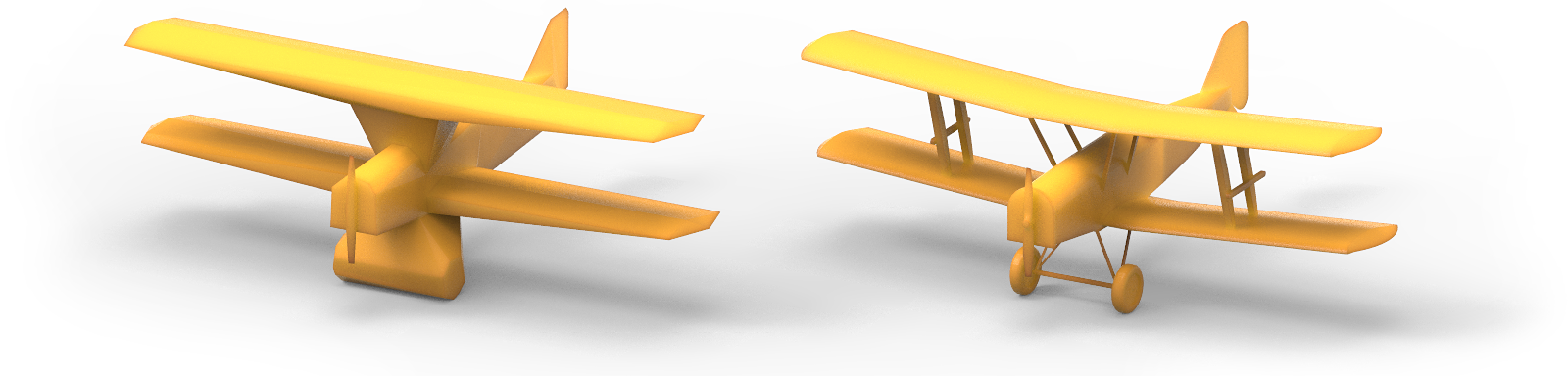}
	\caption{Topology and shape complexity. Complexity of shapes is often related to essential design requirements. For example, the genus zero shape (left) is straightforward to process with most existing graphics algorithms, however, it does not faithfully represents the actual topologically complex shape (right). }
	\label{fig:airplane}
\end{figure}

In order to utilize existing visual content, it is of great importance to automate the analysis and processing of both existing and captured data. A major challenge still facing the fields of computer vision and graphics is the analysis of complex shapes, more often are those of man-made origin. . When the topology of a shape exceeds that of a simple curve or a sphere, many shape analysis or processing algorithms are no longer applicable. For example, contour-based methods often only utilize the outer most boundary and omit the analysis of inner holes or disconnected components. Complications stemming from non-trivial topology of a 3D shape are typically resolved by simplification of the shape to that of sphere or disk topology. However, the complexity in shape topology is often an intrinsic property for a class of shapes (see Figure \ref{fig:airplane}).

In this report we investigate existing works on the topic of modeling and analysis of topologically complex 3D shapes. In the first chapter we survey methods for visual content creation by utilizing existing examples via entire or partial blending. Our focus will mainly be on man-made 3D shapes, especially those having widely different structure and topology. In the second chapter we discuss the correspondence problem, more specifically the mapping of very different shapes which is an essential step for blending methods. In the final chapter we look at different shape representations used in shape creation of complex 3D models. We will survey different tools that allow change in shape topology including implicit surfaces, sculpting tools, and other topology-varying methods.

	\section{Shape Modeling by Blending of Examples} \label{chap:shape-creation}

For novice users who want to prototype new shape ideas, it is much easier to modify on existing shapes than to start from scratch. Users can start the creative process by retrieving existing visual content with the hope of drawing inspiration. Mixing different pieces or styles often yields interesting and sometimes creative concepts, however, simply swapping components of a shape is quite limiting. A better utilization that yields many variations is in continuously blending between different shapes akin to the mixing of primary colors to create a versatile color palette. The majority of shape blending or interpolation have been focused on applications in animation where the smooth transition between frames is a desirable property. In the context of shape creation, there are relatively few successful methods that consider the continuous blending of the whole shape due to the difficulty of mapping and interpolating topologically different shapes. The more common approach to blending for shape creation has been the mixing of whole parts from different shapes with minor modifications. 

In this chapter we survey methods related to shape creation from existing visual elements. Our focus will be on methods that would work on blending of man-made objects. Later in this chapter we survey methods that mainly work on 3D shapes that exhibit large differences in their overall shape and can have very different topologies.

\subsection{Shape Averaging} \label{sec:shape-average}

\begin{figure}[t!]
	\centering
	\includegraphics[width=0.99\linewidth]{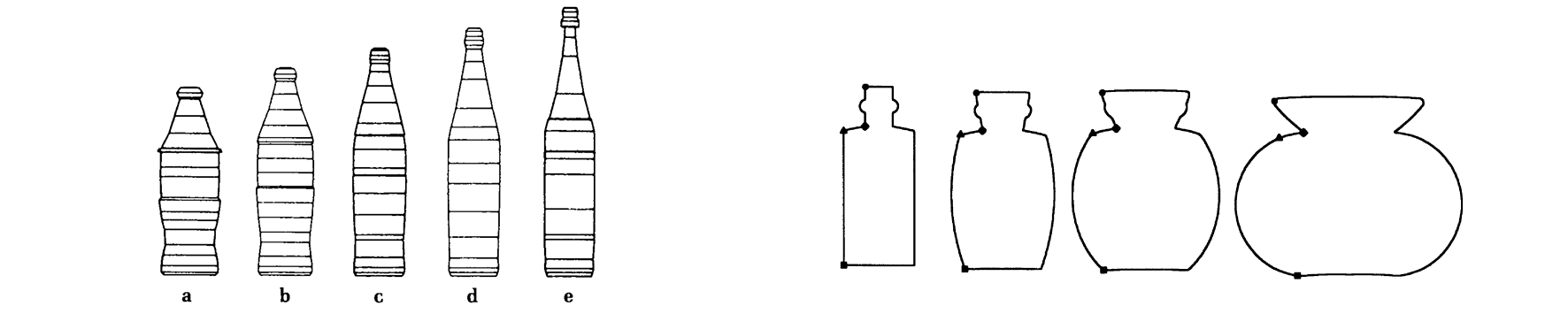}
	\caption{Shape averaging. A set of interpolated and extrapolated results of a Coke bottle from \protect\cite{Chen1989} (left). An example of blending two bottles where correspondence is decided based on shape features \protect\cite{Hui1998} (right).}
	\label{fig:shape-averaging}
\end{figure}

One of the earlier approaches to shape creation in the context of industrial design is the \emph{shape averaging} method which considers whole shapes \cite{Chen1989}. The three stated goals for this work are to help designers predict trends in shape sets, create novel shapes, and stimulate new ideas. While the results presented are rather basic, their pipeline encompasses the major components still in use in more recent shape blending methods; namely, an appropriate shape representation and a correspondence scheme. In this work planar polygons are used as the shape representation to blend both 2D and 3D shapes, the later being approximated by slicing the shape into several contours. Finding a meaningful correspondence between the input shapes is a major part of the process (see Figure \ref{fig:shape-averaging}). In their method they define a polygon vertex correspondence that maps shapes, based on minimum distance, and refine the input as necessary. Later efforts \cite{Hui1998} recognized the need for a better correspondence approach and proposed a feature-based method to match regions that are more reasonably ``blendable''.

While conceptually the averaging (or blending) of entire shapes can be a powerful tool, current tools available in design software focus mostly on the blending of parts of shapes individually. Tools such as the "Blend tool" in Adobe Illustrator, Inkscape's "Interpolate" tool, or the "TweenSurfaces" command in Rhino provide users with the ability to generate in-betweens of two given shape segments. However, all such tools perform low-level shape editing operations that still require a lot of practice and manual effort which prohibits their use by novice users. Furthermore, strong assumptions are implicitly made about the compatibility of blended shapes where an exact mapping is needed which in turn limits the ability to blend shapes of different topology. Finally, we note that the averaging of entire shapes at the same time does not necessarily produce as much of interesting variety in the generated forms as that of blending of individual shape parts. This is the case since the number of combinations for the later approach is naturally larger.

\subsection{Modeling by Example} \label{sec:modeling-by-example}

\begin{figure}[t!]
	\centering
	\includegraphics[width=0.99\linewidth]{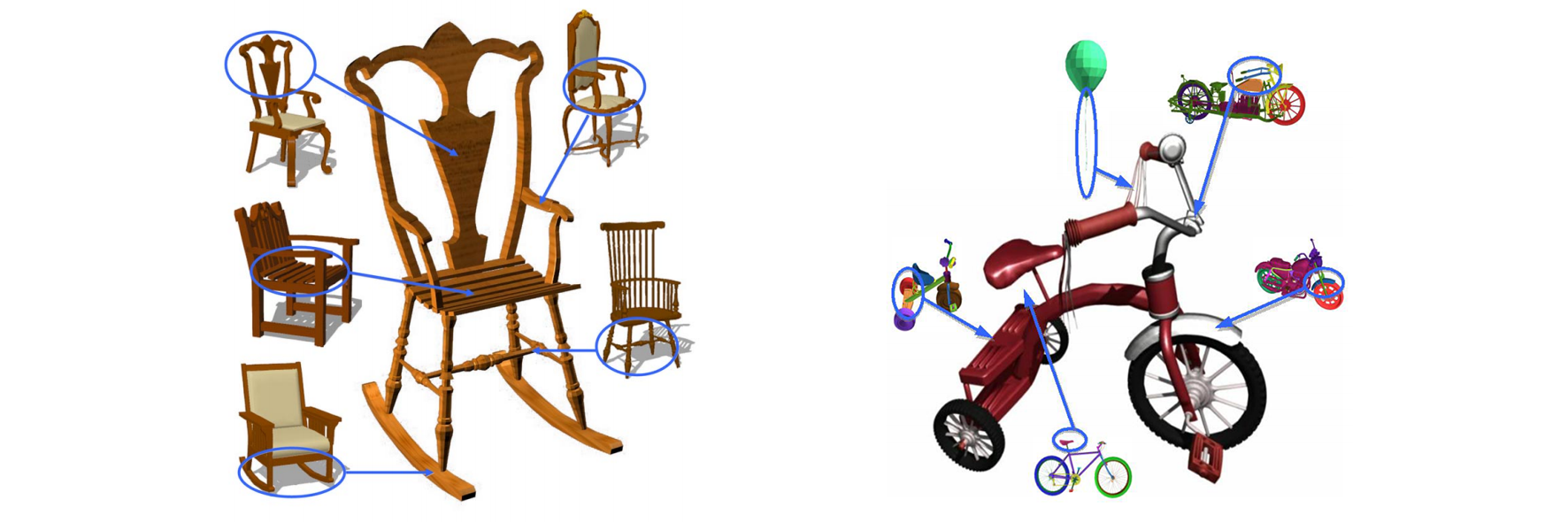}
	\caption{Modeling by example. Parts extracted from different existing 3D models are combined with the help of the user to create new shapes \protect\cite{Funkhouser04}.}
	\label{fig:modeling-by-example}
\end{figure}

The work of Funkhouser et al. \cite{Funkhouser04} represents one of the first comprehensive systems that allowed for easy shape creation by combining parts from an existing dataset (see Figure \ref{fig:modeling-by-example}). The main goal in this work is to develop a 3D modeling tool that requires very little training and user effort making creative shape creation more attainable by novice users. The system simplifies several modeling tasks including part extraction, retrieval, placement and alignment, and stitching of disconnected components. 

The interactions performed during a modeling session include selecting a starting shape by visual inspection or keyword search, then searching the database for alternatives to its parts, and finally applying local edits that identify and stitch the new parts to the active model. The system simplifies part cutting by finding an optimal cut, where cuts along natural seams of the shape are preferred, from a set of user drawn strokes on the shape. The system uses an efficient shape retrieval method, based on an approximate surface distance measure, to query the closest shapes to a query shape from the dataset. Once the user is satisfied with both the cuts of the original parts and the found replacement part, the system computes an automatic placement and alignment of this new part using an efficient variant of the ICP algorithm. The final step is generating a smooth patch to stitch between any open boundaries of the existing shape and the new part which creates a more natural looking model.

Since the introduction of the modeling by example system \cite{Funkhouser04}, many extensions have been proposed that improved on the part segmentation or retrieval process. The work of \cite{Kreavoy2007} developed a compatible segmentation algorithm that tries to identify and extract interchangeable components in order to simplify the shuffling of parts between different models. A more recent example of improvements on retrieval is a proposed 3D shape creation system \cite{Xie2013} that utilizes rough user sketches for the retrieval and replacement of the shape's parts. The method also achieves better style consistency by the addition of contextual information of all pre-analysed parts of the dataset.

While all these methods are much easier to use than modeling from scratch, they still expect some level of manual effort from the user per blended shape. Modeling for a collection of different shapes would entail many queries, either by searching or sketching, in order to create a new set. In addition, they are more effective as tools to model a preconceived design rather than tools to help discover new and creative shape ideas.

\subsection{Data-driven Creative Shape Creation} \label{sec:suggestions}

\begin{figure}[t!]
	\centering
	\includegraphics[width=0.99\linewidth]{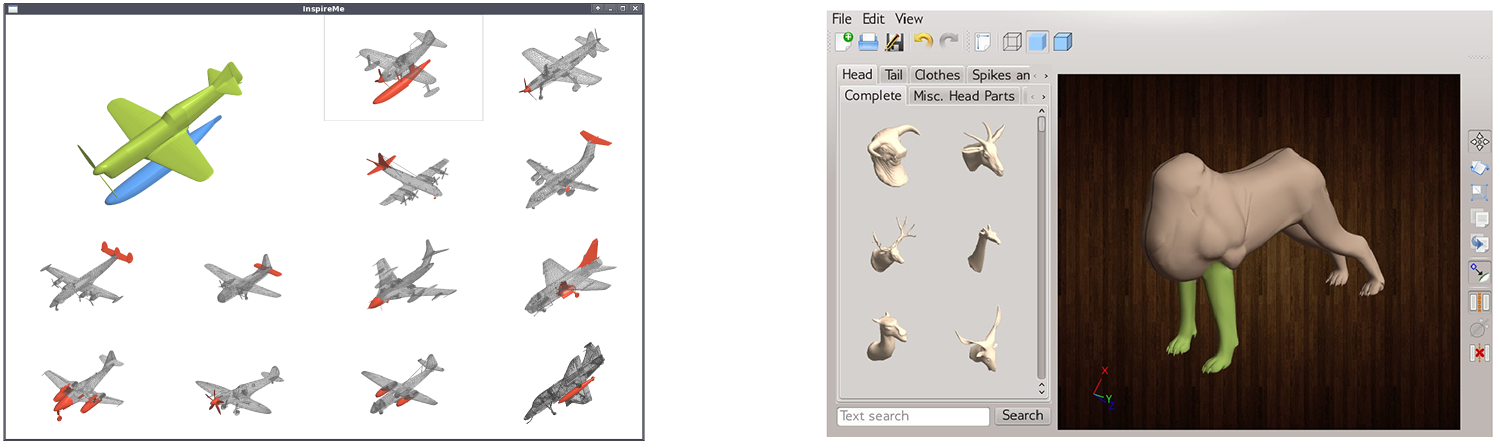}
	\caption{Data-driven creative shape creation. The \emph{InspireMe} system (left) shows the user different suggestions (orange) for the current shape (green) \protect\cite{Chaudhuri2010}. Learning the probability of what parts are better together allows for more smarter suggestions \protect\cite{Chaudhuri2011}. }
	\label{fig:suggestions}
\end{figure}

Research have shown that a valuable source of inspiration is in having many \emph{plausible} existing examples displayed to the user to draw on during stages of design \cite{Chaudhuri2010}. With the increased availability of many different datasets, data-driven approaches are becoming very attractive in the field of computational creativity (see Figure \ref{fig:suggestions}). 

The system introduced in \cite{Chaudhuri2010} automatically provides plausible suggestions of different parts to the user during an iterative modeling session. Starting from a rough initial shape, the system displays multiple suggestions of what parts are likely to be added next. The suggestions are made by an approximate shape retrieval process that compares shape signatures of the current shape against the database of existing shapes. The retrieval process also incorporates the Maximal Marginal Relevance \cite{Carbonell1998} criterion in order to ensure the diversity of the retrieved shapes. Having diverse suggestions is key in helping users create creative and unexpected results. Once the user selects a suggestion, it is placed at an initial position and the user can then edit and modify its transformation. The final step at each part insertion iteration is a basic gluing process that conforms the geometry of the suggestion to the active shape. The system uses a combination of two segmentation algorithms, the shape diameter function (SDF) \cite{Shapira2008} and approximate convex decomposition (ACD) \cite{Lien2008}, to identify the different parts in each model in the dataset. While this system provides a more useful framework for creating blended shapes by proactively suggesting parts, it does not take into account semantic and stylistic properties between the active shape and the shapes considered for suggestion. 

A later system \cite{Chaudhuri2011} was proposed with the goal of better suggesting semantically and stylistically more relevant parts during assembly. The system depends on learning a probabilistic graphical model (Bayesian network) trained on an analyzed shapes database. A segmented and consistently labeled dataset is required for each class in the database. This preprocessing step is done semi-automatically with the help of the supervised learning method in \cite{Kalogerakis10}. The set of segmented components is then clustered into same style sets based on geometric features including shape diameter \cite{Shapira2008}, curvature, shape context, and other surface-based features. The probabilistic model then trains on the resulting part labels and style clusters and learns dependencies between labels, part adjacencies, number of parts, and symmetries. Probabilistic inference is then performed at each assembly step to generate a list of ranked parts. The ranking is based on compatibility with the existing parts on the active shape. While the probabilistic model helps suggest more meaningful parts to blend together, it is not sufficient for generating entire plausible blended shapes automatically.

The work of Kalogerakis et al. \cite{Kalogerakis2012} improves on the probabilistic model of the earlier system. The new model allows the system to automatically generate a large number of plausible blended shapes from a small set of example shapes. The key idea is in relating probabilistic relationships, between geometric and semantic properties of parts, to learned latent causes of structural variability especially at the level of the entire shape. For a given set of compatibly segmented shapes, the model learns a probabilistic model that accounts for cardinality of a certain part type, its geometric features, and adjacency relations between the different part types. The learning process is done offline and the resulting learned model is compact and ready for synthesis by sampling the model for all parts needed for a plausible shape. The placement of the selected parts is further optimized by considering symmetric relations of parts and by minimizing disconnections and relative size differences. The system was shown to be able to expand the original example set with new plausible models by an order of magnitude. The complexity and variability of the generated models is dependent on the given input. 

Since parts are picked to match an existing model that is trained from a fixed set, it is unlikely that unexpected inspiring shapes are generated. In order to generate creative blended shapes, the user still needs to be the creative driving force by adding randomness to the part or creating unique configurations that never existed in the original database.

\subsection{Set Evolution for Creative Modeling}

\begin{figure}[t!]
	\centering
	\includegraphics[width=0.99\linewidth]{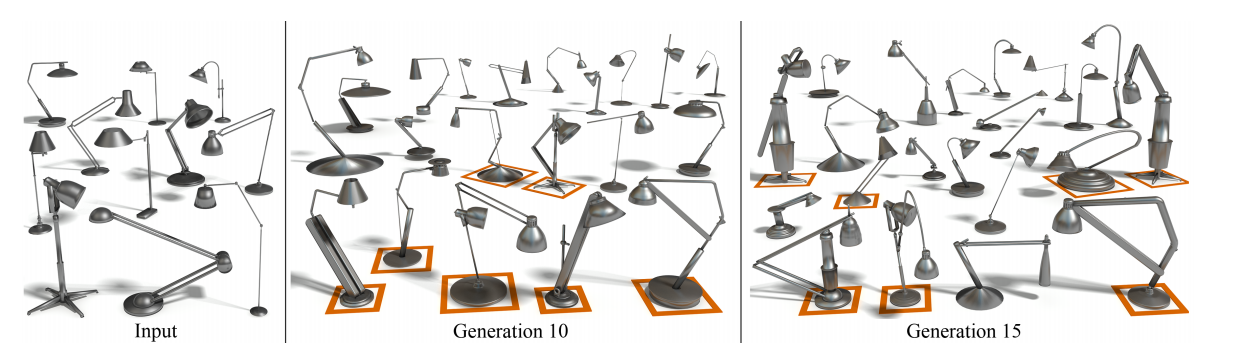}
	\caption{Set evolution for creative shape modeling. A small set of input shapes evolve into inspiring variations using part crossover and mutation operations that are evaluated at each generation as a set by the user \protect\cite{Xu2012}.}
	\label{fig:set-evolution}
\end{figure}

Blending by part assembly can generate a large set of new shapes, but an even larger set is one where groups of parts or individual components themselves undergo geometric and topological transformations. Inspired by principles from the theory of evolution, the work of Xu et al. \cite{Xu2012} takes a population of shapes and work on breeding new generations of diverse shapes. The system evolves shape sets by performing part crossovers and mutations and evaluate the fitnesses of a \emph{set} of shapes based on the user's preference. The diversity of the generated shapes is essential for inspiring creative new designs. In each generation, the user is shown a new set of evolved shapes and are expected to evaluate their preference of some shapes. This evaluation translates into a fitness score that guides the process of generating subsequent generations (see Figure \ref{fig:set-evolution}).

The input dataset is assumed to be pre-analyzed, where relations such as symmetry and proximity between the parts are stored. The parts are enclosed by deformation controllers such as bounding cuboids and generalized cylinders as described in the deformation model of \cite{Zheng11}. Group of parts that represent a specific component of the shape's class, such as the backs of chairs, are identified and corresponded. A crossover process exchanges parts of components between parent shapes resulting in a new offspring. With a specified probability, some parts undergo a mutation which involves mainly a limited random scaling of the part's controller. Stored symmetry and proximity constraints between different controllers are enforced at each step ensuring a well connected and symmetric offspring. 

Set evolution for creative shape modeling shows the potential of how fine-grained blending of shape parts produces interesting and possibly inspiring results. By having both the crossover and mutation of parts, the generated shapes exhibit large variability on both the structure and geometry compared to shapes in the database.

\subsection{Recombination of Unlabeled Parts}

\begin{figure}[t!]
	\centering
	\includegraphics[width=0.99\linewidth]{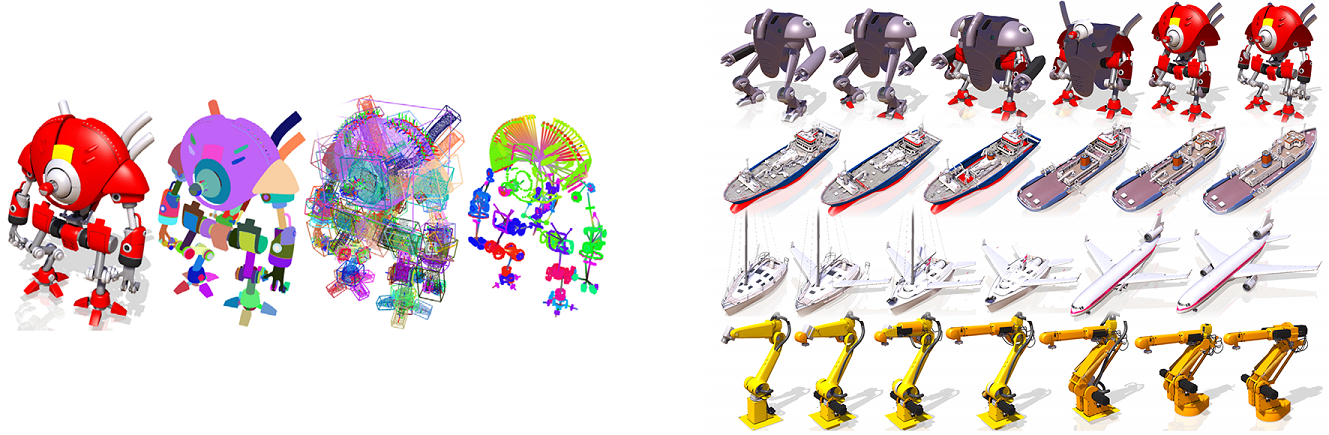}
	\caption{Blending via recombination of unlabeled parts \protect\cite{Jain2012}. After analyzing symmetry and contact relations between the different shape parts (left), many shape combinations are immediately available as a blending process between two shapes (right). }
	\label{fig:recombination}
\end{figure}

The methods described earlier leverage large sets when generating new shapes. A simplified systems by Jain et al. \cite{Jain2012} that blends between only two input 3D shapes of varying complexity. The system assumes that the parts in the input shapes are unlabeled. Each input shape is segmented and analyzed to encode part contacts, symmetry relations between parts, and a coarse-to-fine shape hierarchy. When creating new blended shapes the system starts with a shape matching step, then blending of shape parts, followed by contact enforcement (see Figure \ref{fig:recombination} for results). 

The system performs hierarchical matching of the input shape pairs by trying to construct for the target shape the best matching hierarchy from the source shape. The goal is to find a one-to-one mapping between the nodes of source and target hierarchies that will be used in the blending step. Starting from the coarse levels the algorithm aligns the parent nodes and assigns, with respect to source hierarchy, child nodes based on the closest relative position inside the parent source node. Any remaining disconnected nodes in the target shape are forced to be connected by modifying the hierarchy of the target.

Blending of the shapes is controlled by a weight parameter where matched nodes at the finest level in the hierarchy are progressively replaced. The priority of the parts to replace during a blend from source to target is determined by the combined size of the source and target nodes. The blending of multiple shapes can be achieved by taking two intermediate results as source and target and applying the blending step. During the blend, parts replaced are often disproportionation leading to disconnections of the shape. The system solves this issue with a contact enforcement process using a simple mass-spring system. Masses are created at a node ands its contact points with the neighbors. One set of springs are generated between these masses, within a node, and another between the different contact points of connected nodes (with zero length). When parts are replaced during a blend, the system pulls the different nodes to enforce these spring constraints resulting in a well connected shape.

This system greatly simplifies the blending of a pair of shapes of complex geometry with no assumption on the structure. Since the shape analysis is purely geometric, issues on plausibility are bound to be present. The matching of parts does not take into account functional similarity of the matched parts which sometimes result in either missed functionality or redundancies. Furthermore, input shapes do not necessarily have all their parts being compatible, i.e., parts on the source shape might have no equivalent on the target and vice versa.

\subsection{Functionally Plausible Shape Combinations}

\begin{figure}[t!]
	\centering
	\includegraphics[width=0.99\linewidth]{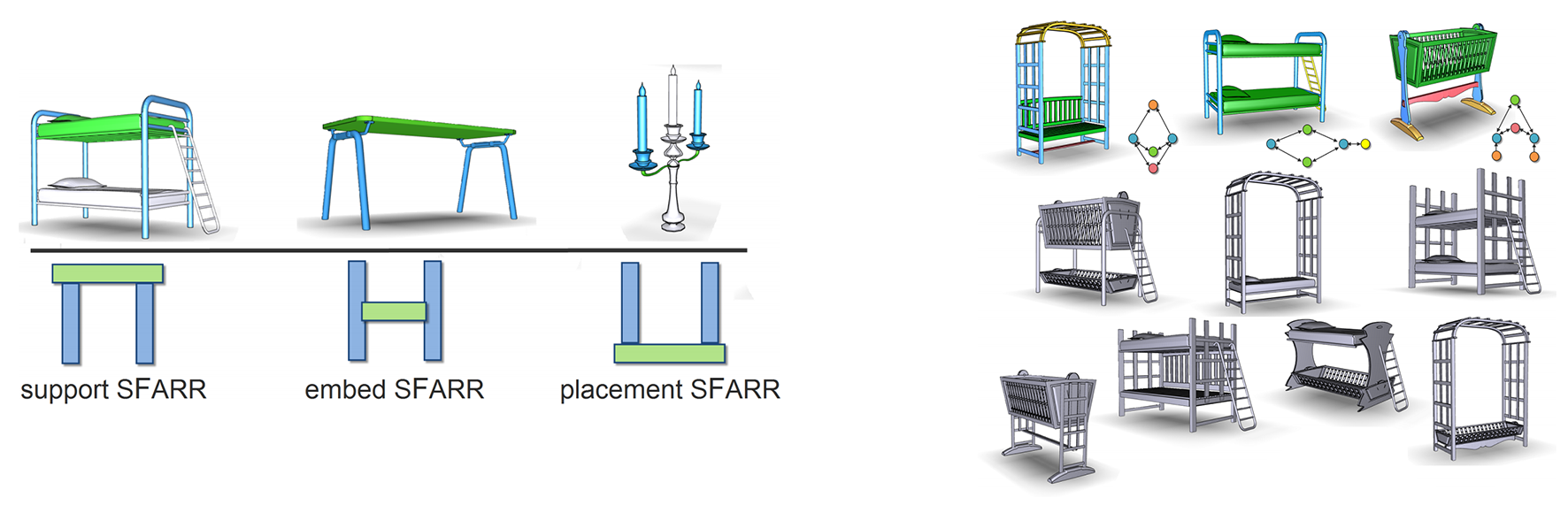}
	\caption{Functionally plausible shape combinations \protect\cite{Zheng2013}. By extracting a specific arrangement of symmetric parts (left), functional sub-structure can be reasonably interchanged among different shapes in the set (right).  }
	\label{fig:functional}
\end{figure}

The resulting shapes when blending using methods that require labeled shape collections often appear plausible; on the other hand, methods that relay solely on geometric features to identify replaceable parts typically generate implausible results. This is especially apparent with the loss of functionally for man-made shapes when different functional parts are exchanged. The system by Zheng et al. \cite{Zheng2013} proposes a geometric approach to replace a collection of parts together in order to ensure preserved functionality. The key idea is that many man-made shapes share similar arrangements of parts that serve the same function, thus they are compatible and are good candidates for replacement among the dataset. The system analyzes the shape's structure to identify substructures having several symmetric parts that are connected with a shared base part. Such sub-structures reflect three types of functional arrangements including \emph{support}, where the symmetric parts connect underneath the base part, \emph{embed}, where the base connects at the side of the symmetric parts, and \emph{placement}, in which the base supports the parts with respect to the upright direction (see Figure \ref{fig:functional}). When synthesizing new combinations, the compatible arrangements are shuffled and correctly placed then deformed to fit with the remainder of the shape. Relating geometric substructures to shape functionality shows some promise in better ensuring the plausibility of shapes resulting from blended examples. This is especially advantageous when the dataset lacks semantic labels. However, having strict requirements on what parts are replaceable has the disadvantage of having the user miss out on some other creative part arrangements.

\subsection{Continuous Blending of Arbitrary Shapes}

Modeling typically involves the manipulation of shape geometry either by modifications of existing elements, e.g., lines or triangles, or by the insertion of such elements. Recent modeling systems combine or blend from example shapes by completely replacing existing parts of the shape, i.e., they perform a discrete replacement rather than a continuous blend. A more capable solution would be to apply a continuous transformation that interpolates between the shape of a part from the source and its corresponding part on the target (or several targets). 

\subsection{Image morphing}
\begin{figure}[t!]
	\centering
	\includegraphics[width=0.99\linewidth]{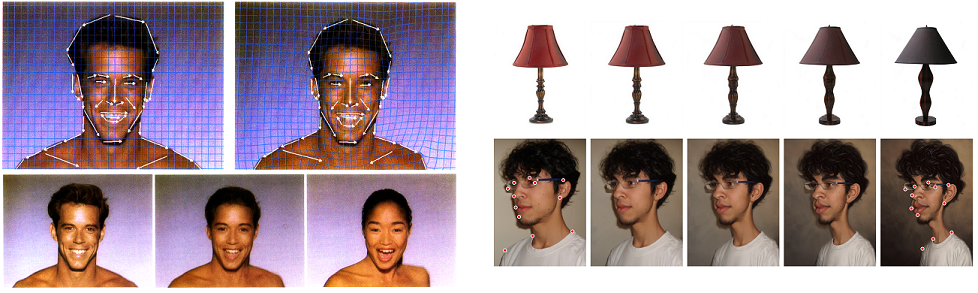}
	\caption{Image morphing. A famous example of image morphing use in media \protect\cite{Beier1992}. State-of-the-art in image morphing produces quality morphs while requiring minimal user input \protect\cite{Liao2014}. }
	\label{fig:image-morphing}
\end{figure}

Continuous blending have been successfully used by the video and film industry when creating visual effects. Earlier efforts have been focused on \emph{image morphing} to achieve natural transitions between images with varying colors and structure \cite{Smithe1990,Wolberg1990}. The general framework involves warping the input images with respect to certain features and applying color interpolation \cite{Beier1992,Wolberg1998,Gomes1998}. While this framework for image blending have remained mostly the same, more recent methods have made advancement on simplifying blending by finding good warping with very few user-drawn points on input images \cite{Liao2014} or video frames \cite{Liao2014b} (see Figure \ref{fig:image-morphing}). The in-betweens resulting from image morphing are rarely useful as independent creations as they often suffer from visual artifices including ghosting, mismatched colors, or disconnected structures. While these issues are overlooked in applications relating to animation they are not acceptable in the context of shape creation.

\subsubsection{2D Polygon Morphing}
The blending of shapes represented by \emph{piecewise linear} elements have been shown in shape averaging \cite{Chen1989} and 2D shape blending \cite{Sederberg92} (see Figure \ref{fig:basic-shape-morphing}). These methods work on transforming geometric elements, most commonly the vertices, of a source and target shape. The morphing process, whether for 2D or 3D shapes, generally involves finding a correspondence, that matches the vertices of source and target, and assigning an interpolation scheme that defines how intermediate results are computed \cite{Lazarus98,Alexa02}. Different attempts have been made to automatically compute the best correspondence between source and target shapes. The physically based approach proposed in \cite{Sederberg92} works on minimizing the work required to bend and stretch one 2D shape into the other. The later approach of \cite{Liu04} solves the correspondence problem by matching the most prominent perceptual feature points on the input shapes. 

\subsubsection{3D Mesh Morphing}
\begin{figure}[t!]
	\centering
	\includegraphics[width=0.99\linewidth]{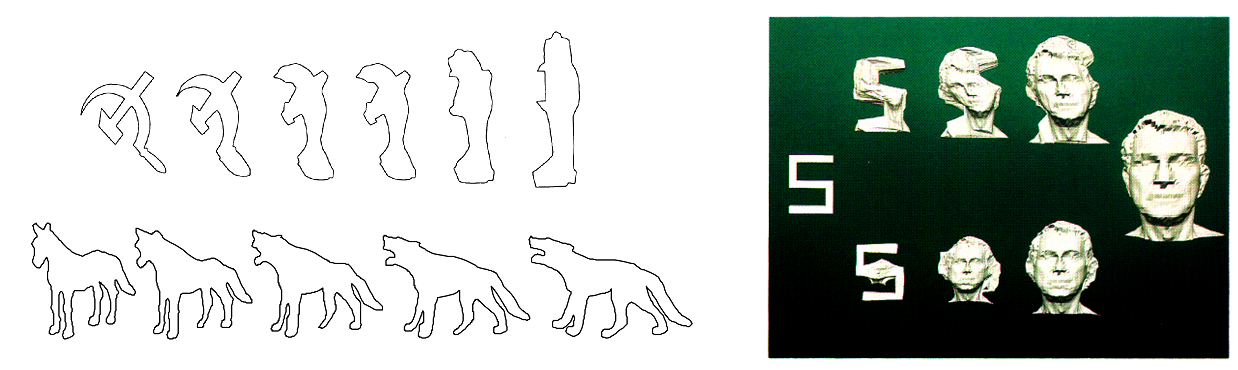}
	\caption{Basic shape morphing. For 2D polygon morphing (left) different mechanisms for correspondence include physically-based (top) \protect\cite{Sederberg92} and perceptually-based (bottom) \protect\cite{Liu04} approaches. Early concepts in 3D mesh morphing \protect\cite{Kent1992} have not changed significantly. }
	\label{fig:basic-shape-morphing}
\end{figure}

In the early work of Kent et al. \cite{Kent1992} the correspondence between two 3D meshes is computed by projecting the shapes onto a unit sphere (see Figure \ref{fig:basic-shape-morphing}). Mapping the source and target to a common domain allows for the correspondence of shapes having different mesh connectivity. Other methods suggested the automatic embedding of input meshes onto a desk while minimizing angle distortion cased by the embedding \cite{Kanai97,Kanai2000}. Multi-resolution approaches, such as \cite{Lee1999}, simplify the correspondence problem by considering a few user defined corresponded points then working on computed coarse base domains that are easier to solve the correspondence problem for. A more efficient multi-resolution method \cite{Michikawa2001} used subdivision fitting to create a common base mesh and was shown to be useful for real-time interpolation of multiple meshes. The use of common base meshes was also explored in \cite{Praun2001} where consistent mesh parameterizations are established allowing for n-way shape blending. A feature-preserving parametrization algorithm proposed in \cite{Kraevoy2004} reduced the distortions caused by the mapping to base meshes as computed by earlier methods. As in the 2D case, the quality of most surface-based 3D morphing methods also depends on providing a reasonable correspondence of feature points on the input shapes. In \cite{Athanasiadis2012} an automatic feature-based morphing method uses pattern matching to identify feature points for 3D morphing without any user input. Most of these methods assume that the input shapes are of the same topology as dictated by the mapping strategy. Perhaps another method for mesh morphing worth noting is the skeleton-based approach in \cite{Blanding2000} that uses the medial axis as its underlying representation. While it is able to morph between shapes of varying topology, it has not been adopted to handle shapes more complex than the ones shown in their paper. This is most likely due to its dependency on both accurate and clean skeletons for the reconstruction of the interpolated in-between; both a computational heavy and unstable process.

\subsubsection{Topology Varying Morphing}
Both 2D and 3D shape blending methods mentioned above are restricted to shapes having similar topology. While mainly designed for animation and visual effects, some of these morphing methods have been tried in applications relating to product design with minor success \cite{Chen2003,Nashvili2005,Kang2010,Li2011}. The fact that most man-made objects have complex topological structures prohibits the use of these approaches in general shape modeling. Nonetheless, several \emph{geometry-based} and \emph{volume-based} methods that support some topological changes have been proposed especially for the application of animation \cite{Lazarus98,Alexa02}.

\subsubsection{Geometry-based Morphing}

\begin{figure}[t!]
	\centering
	\includegraphics[width=0.99\linewidth]{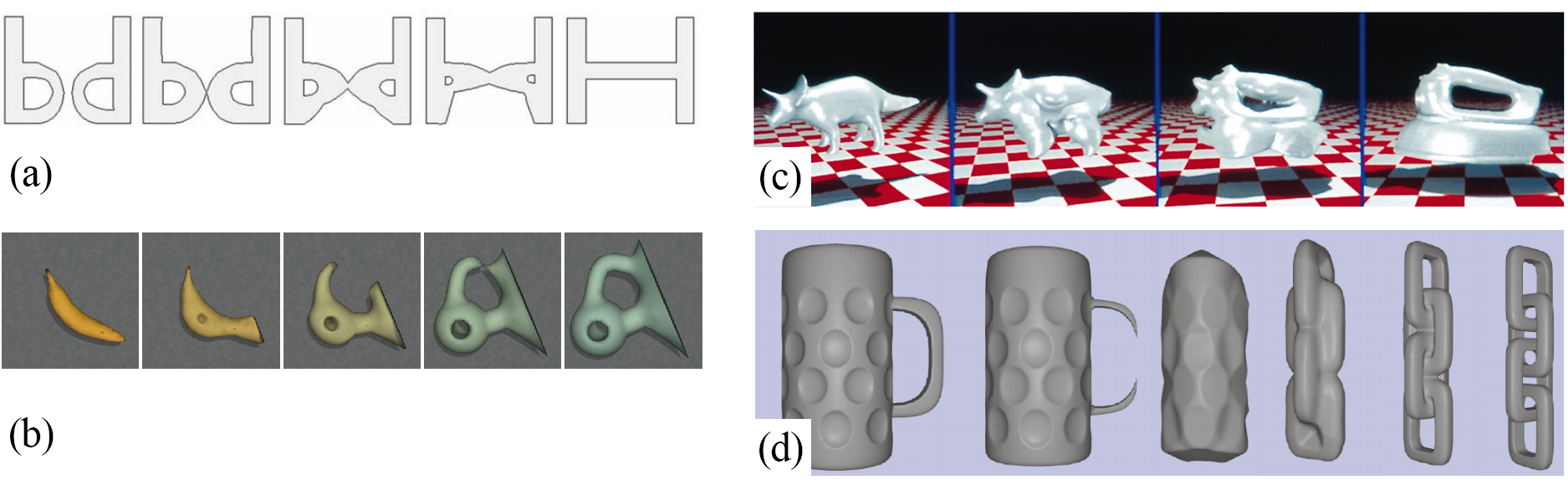}
	\caption{Topology varying morphing. (a) Natural morphing between 2D polygons having different topology with the help of user input \protect\cite{Liu2005}. (b) Early work on morphing 3D meshes of different topology required exact specification of how surfaces split or merge during the blend \protect\cite{DeCarlo96}. (c) The use of distance fields allows the freedom of morphing shapes of any topology; some control is possible using warping functions \protect\cite{Cohen1998}. (d) The level-set approach also allows morphing between any two topologies while ensuring intermediate shapes remain connected \protect\cite{Breen2001}.  }
	\label{fig:topology-shape-morphing}
\end{figure}

The geometry-based morphing framework introduced in \cite{DeCarlo96} allows for the blending of topologically different 3D meshes (see Figure \ref{fig:topology-shape-morphing}(b)). This method, however, requires substantial manual effort from the user for assigning the correspondence on the input and providing the timing of topology changing operations. Later effort presented in \cite{Surazhsky2001} proposed a solution using a modified reconstruction algorithm for 2D polygon morphing, however, the resulting in-betweens were not very smooth or natural looking. An automatic 2D polygon-based blending method described in \cite{Liu2005} produced better morph shapes using a set of topological operations which are also controllable by the user (see Figure \ref{fig:topology-shape-morphing}(a)). While this method produced reasonable morphs in 2D, it is not immediately clear how the different algorithm components are transferable to 3D surfaces. A suggested 3D method shown in \cite{Lee2006} simplifies handle regions on the input 3D shapes using hole filling procedures to enforce topological isomorphism to a sphere which enables the use of traditional mesh morphing methods on shapes of high genus. Finally, a notable work to mention is the system described in \cite{Zhao2003} for interactive component-based mesh morphing which allows, to some degree, shapes having different number of parts to morph. The system works on shapes segmented into compatible parts and allows for parts to correspond to a null-component which effectively shrinks (or grows) extra parts on either shapes into (or from) degenerate geometry. Geometry-based methods are generally limited by their ability to handle complex topology, which may explain their limited adoption in the blending of shapes as a modeling process. 

\subsubsection{Volume-based Morphing}

A key advantages of volume-based morphing is that it allows shapes of any topology to be blended together. The early work of Hughes \cite{Hughes1992} demonstrates the capability and simplicity of volume morphing by interpolating the Fourier transforms of the input shapes. A similar method by He et al. \cite{He1994} apply volume morphing in the wavelet domain having the advantage of establishing a correspondence at a low level. The work of Cohen et al. \cite{Cohen1996} introduced morphing based on Distance Field Interpolation (DFI) \cite{Levin1986} which is able to interpolate between 2D contours of any topology. The morphing is combined with a warping process to better align features of the input shapes. The method was later extended to 3D volumes in \cite{Cohen1998} where the warping function was computed for the volume representations (see Figure \ref{fig:topology-shape-morphing}(c)). Volumetric shape blending was also done by interpolating implicits \cite{Savchenko1995} and was shown to produce smooth morphing of shapes with any topology in both 2D and 3D \cite{Turk1999,Pasko2004}. More strict user control of volumetric morphing was the key contribution in Weng et al. \cite{Weng2013}. In their method, sparse user specified correspondences were used to solve an optimal mass transport problem that enforced local rigidity during the morph unlike early DFI methods. 

Volume morphing without any required user input or warping was described in the level-set morphing method of \cite{Breen2001} (see Figure \ref{fig:topology-shape-morphing}(d)). This approach interpolates between two shapes by deforming one to exactly match the other in a surface evolving process. The evolution from source to target is driven by a signed distance transform from source volume to target volume. Besides being able to blend shapes of any topology, the method also has the advantage that blended in-betweens remain connected throughout the blend, i.e., there are no sudden materialization of the target shape as in DFI methods.

Volume-based morphing is a powerful mechanism for generating continuous blends of arbitrary shapes. While achieving wide use and success in visual effects \cite{Houston2006,Museth2013}, it has yet to be used for generating blending of shapes containing multiple semantic parts, e.g., man-made objects exhibiting complex topology. Such cases would require solving the problem of dense correspondences for shapes with varying topology in order to produce well connected and feature persevering results.

	\section{Correspondence of Complex 3D Shapes} \label{chap:correspondence}

In shape analysis, the problem of correspondence, or matching, is a fundamental and challenging task \cite{vanKaick_survey_10,Chang2011}. Proposed solutions for finding a \emph{global} meaningful correspondences are mostly applicable to shapes having very similar structure and topology. A solution to the more difficult problem of \emph{partial} correspondence might help match topologically very different shapes by finding the best partial matches of their parts. In general, the desired type and quality of a shape matching is dependent on the application it is needed for. In applications such as shape interpolation and style transfer, the correspondence is mostly needed at the part level. 

While most organic shapes have a fairly clear one-to-one correspondences, human-designed objects often exhibit large variability in topology (see Figure \ref{fig:airplane}). The complexity in topology for man-made objects are often a result of assembilibity, styling and aesthetics, affordance of certain functionality \cite{Norman2013}, or optimal use of material \cite{Bendsoe2003}. For example, an assemblable object often splits up into multiple parts having the same functionality, such as table legs, to facilitate ease of shipping. Another example is having holes in a single functional part of a bicycle which allows for the same functionality with less material, thus, making the bicycle lighter.

In this chapter we discuss some major methods for 3D shape correspondences of very different shapes. We start by first describing the general trends for computing dense correspondences of topologically similar shapes. Next, we discuss possible solutions for the correspondence of topologically different shapes via correspondence of a sparse set of feature points.

\subsection{3D Dense Shape Correspondence}

\begin{figure}[t!]
	\centering
	\includegraphics[width=0.99\linewidth]{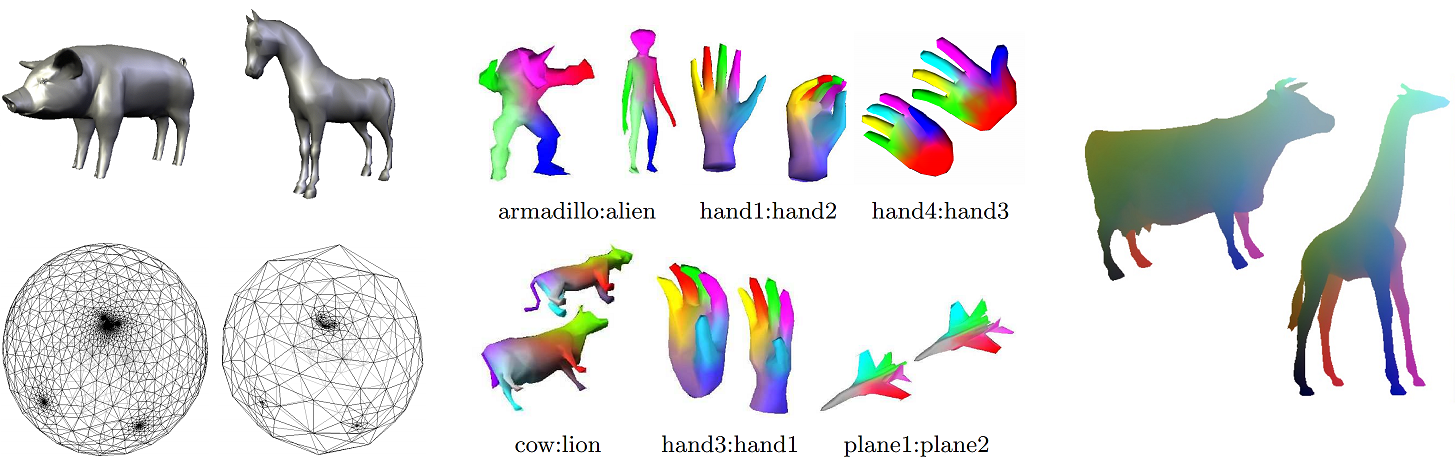}
	\caption{Dense shape correspondence. (left) Early methods of surface correspondence map the two surfaces to a common domain while matching prominent shape features \protect\cite{Alexa02}. (middle) Spectral methods embed shapes into comparable spaces which minimizes shape disparity under rigid transformations \protect\cite{jain2006robust}. (right) The blended intrinsic maps method \protect\cite{Kim11} uses a combination of nearly isometric maps to automatically correspond genus zero shapes. }
	\label{fig:dense-correspondence}
\end{figure}

Common approaches for computing dense correspondence between two shapes map both to a common domain, such as a sphere or a coarse base mesh, while aligning some of their feature points \cite{Alexa02,Kraevoy2004,Athanasiadis2012} (Figure \ref{fig:dense-correspondence}(left)). Other approaches normalize the shapes by embedding them in a comparable space using multidimensional scaling (MDS) \cite{Elad03,Bronstein08} or the spectral method \cite{Zhang2010} (Figure \ref{fig:dense-correspondence}(middle)). Later methods for surface mapping solve the dense correspondence problem by first finding the best matching for a small set of feature points, under near-isometric transformations, then extending the result to the entire surface \cite{Lipman09,Kim11}. More specifically, the Blended Intrinsic Maps (BIM) approach \cite{Kim11} has since become a baseline for point-to-point correspondences (Figure \ref{fig:dense-correspondence}(right)). Rather than corresponding points on the shape, the method of \cite{Ovsjanikov2012} proposed corresponding \emph{real-valued functions} over the shapes resulting in an efficient and accurate shape matching. 

While all these method compute dense point-to-point mapping between shapes, they are not applicable when computing correspondences of topologically different shapes. Attempts to adapt some of these methods in matching shapes having different topology include the work in \cite{Sandilands2014}, where BIM were made to handle the matching of similar shapes with large holes, and the extension of functional maps \cite{Huang2014}, where consistency of the functional maps across a shape set were used in exploration and co-segmentation of shapes with variable topology. 

\subsection{Correspondence via Sparse Features Matching}

The matching between two or more 3D shapes has been widely studied in point cloud registration and similarity-based shape retrieval \cite{vanKaick_survey_10}. The most common techniques for finding partial matchings or when matching dissimilar parts is to simplify the search to that of matching representative elements then evaluate the match with an appropriate objective function. 

\begin{figure}[t!]
	\centering
	\includegraphics[width=0.99\linewidth]{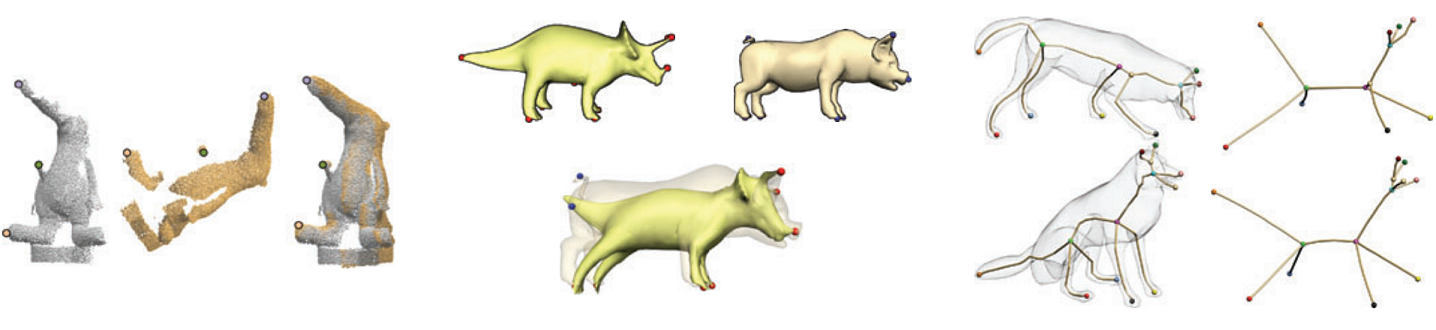}
	\caption{Shape correspondence via sparse features matching. (left) Surface registration using the 4PCS method allows matching of surfaces that align under rigid transformation \protect\cite{Aiger2008}. (middle) The deformation-driven method \protect\cite{Zhang08} automatically finds correspondences of 3D shapes under non-rigid transformations. (right) The method of \protect\cite{Au10} utilizes curve-skeletons in a voting approach for surface correspondence. }
	\label{fig:sparse-correspond}
\end{figure}

For surface registration, especially in the context of raw scans, the underlying assumption is that the surface does not change in shape and so near exact overlapping regions can become the focus of the search. In Li et al. \cite{Li2005multiscale} prominent features are extracted and used to compute approximate alignments of the input surfaces followed by the Iterated Closest Point (ICP) process \cite{Chen1991,Besl1992}. A later method uses the concept of 4-points congruent sets \cite{Aiger2008} to efficiently find the best alignment which can be followed by few ICP steps for further refinement (Figure \ref{fig:sparse-correspond}(left)). While these registration methods are able to align shapes in the presence of large noise and only partial overlap, they are not applicable when large enough changes negate the assumption that the solution is a rigid transformation. Non-rigid registration adds more complexity to the problem since different corresponding regions of the input shapes might deform independently of the others \cite{RegistrationSurvey2013}.

Another strategy to compare between shapes that are articulated or deformed is to represent them as a \emph{graph} of their most prominent connected parts. Using a graph representation significantly reduces the search space and minimizes the effect that deformation or other shape perturbations might have on the correspondence search. For 2D shapes, graph representations were considered for shape matching by looking at their shock graphs \cite{Siddiqi1999,Sebastian01}, which are based on the medial axis \cite{Blum1967}, and searching for the optimal sequence of graph edits that transform one shape into the other. Other representations used to analyze the topology of a shape look at the evolution and arrangement of level sets of functions defined over the shape \cite{Biasotti2008}. The correspondence method presented of Hilaga et al. \cite{Hilaga01} matches Reeb graphs \cite{Reeb1946,Shinagawa1991}, constructed using the integral of the geodesic distance between all surface points, in a multi-resolution fashion. The use of Reeb graphs was later extended in \cite{Biasotti06} for the task of finding sub-part correspondences using graph-matching techniques. Curve skeletons \cite{Au08} were used in a more efficient and robust correspondence technique described in \cite{Au10}. The method utilizes curve skeleton nodes as the shape features in a voting scheme in which the matching with the most votes is selected (Figure \ref{fig:sparse-correspond}(right)). 

Other methods for non-rigid registration and correspondence use global deformation measures resulting from deforming one shape to the other \cite{Huang08,Zhang08}. The deformation-driven shape correspondence of \cite{Zhang08} automatically searches for a sparse point correspondence that results in a minimal-energy surface deformation (Figure \ref{fig:sparse-correspond}(middle)). The method identifies sparse features by looking at shape extremities.

We have seen that with feature correspondence it is possible to match shapes that can be very different geometrically. The voting method in \cite{Au10} can even tolerate small differences in the topology of the matched shapes. Other correspondence methods could also be applied by simplifying the topology of the input shapes by removing holes \cite{Lee2006} or using proxy meshes that enable methods that handle genus zero models \cite{Sandilands2014}. However, most man-made shapes exhibit high topological variability which limits the performance of said methods. Matching such shapes would entail solving the correspondence problem where shape parts might correspond in a one-to-one, one-to-many, or one-to-nothing manner which increases the search space. This topological difference also limits the applicability of current methods that generally relay on assumptions of relative geometric similarity between input shapes. For example, the order in which parts are connected in human or animal models are generally fixed, where as legs and other supporting elements in chair or table models vary significantly.

\subsection{Correspondence and Shape Structure Analysis}

Man-made objects exhibit large variability in both geometry and topology \cite{Mitra13}. Recent methods that work on identifying consistent maps or segmentations of a collection of shapes are looking at leveraging knowledge from a group or incorporating high-level shape semantics. This direction has proven to be successful since man-made objects exhibit more similarity of structure and how the parts are connected than they do in low-level geometric features.

\begin{figure}[t!]
	\centering
	\includegraphics[width=0.99\linewidth]{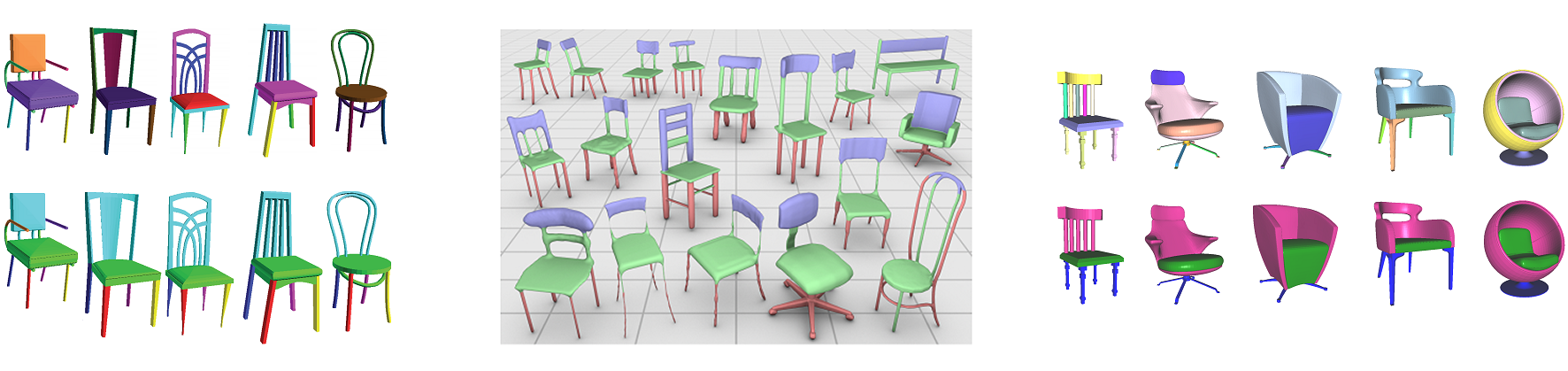}
	\caption{Correspondence and shape structure analysis. (left) Consistent shape segmentation is obtained hierarchical clustering of faces within and across a set \protect\cite{Golovinskiy09}. (middle) Unsupervised co-segmentation by clustering mesh faces in descriptor space \protect\cite{Sidi11}. (right) Consistent segmentation at the part-level by extracting similar part arrangements in a shape set \protect\cite{Zheng14}. }
	\label{fig:part-correspondence}
\end{figure}

In Golovinskiy et al. \cite{Golovinskiy09} a collection of shapes from the same class are analyzed together in order to assign a consistent decomposition of the prominent parts in the shape's class (Figure \ref{fig:part-correspondence}(left)). The method pre-aligns the models and construct a graph where edges connect faces within the model with close-by faces of other models after alignment. Following this graph construction a greedy hierarchical clustering algorithm \cite{Golovinskiy2008} would assign a consistent segmentation of mesh faces across the collection. A major limitation of such methods is that they relay on mesh alignment in handling the variability in shape collections which is not always applicable since not all parts can be brought into alignment with one global rigid transformation. 

Later works followed this trend of leveraging the entire set of shapes when solving for the segmentation and correspondence problem. The work of Kalogerakis et al. \cite{Kalogerakis10} presented a data-driven approach for consistent segmentation that trains a classifier to learn from labeled data the different parts of a shape class. Later work described by Sidi et al. \cite{Sidi11} proposed an unsupervised \emph{co-segmentation} by clustering in the space of shape descriptors (see Figure \ref{fig:part-correspondence}(middle)) instead of the spatial coordinates as in \cite{Golovinskiy09}. The work of Kim et al. \cite{Kim2012} produces a \emph{fuzzy correspondence} by looking at similarity of shapes in the set in an embedded space. Deformable templates were suggested for capturing the variability of man-made 3D shape collections \cite{Ovsjanikov11}. A more comprehensive system in Kim et al. \cite{Kim2013} proposed using an initial template and optimizing for both part segmentation and surface correspondence in order to produce a set of probabilistic part-based templates that reflect the shape collection. The approach of Huang et al. \cite{Huang2014} utilized functional maps across a shape collection which allowed for the correspondence of the parts in different shapes. Most these methods assume a consistency in the geometric properties, such as relative part position and extent, of the parts across the shape set. A recent approach from  Zheng et al. \cite{Zheng14} focused the search on the arrangement of abstractions of shape parts. This allows the method to statistically discover consistent sub-structures rather than individual parts which geometrically can be very different (Figure \ref{fig:part-correspondence}(right)). Another recent work by Tevs et al. \cite{Tevs14} looks at relating shapes by comparing symmetry and regularity relations of the input models.

\subsection{Shape Complexity and Correspondence}

The complexity of the correspondence problem is largely dependent on the type of shapes to match and the amount of similarity to be expected. In the context of scan registration or tracking it is reasonable to search for exact geometry since inputs most likely overlap. For shapes that only exhibit slight deformations, such as articulated animal shapes, feature matching is a viable solution as demonstrated by many methods that consider features at extremities. Complex shapes beyond those exhibit differences in both overall shape topology and the geometry and cardinality of components. The consistent alignment of the shape collection can provide strong clues, however, current proposed methods do not yet allow for more fine grained correspondences including part correspondences of one-to-many or one-to-nothing that are common for topologically different shapes. The recent trend in corresponding shapes by considering the arrangement and structural relations of parts show great promise for corresponding complex man-made shapes.

	\section{Tools for Modeling Shapes of Complex Topology} \label{chap:representations}

Early research in computer graphics \cite{Baumgart74,Weiler86} presented different 3D shape representations, along with construction and editing operations, which are still used in modeling tools today \cite{Maya2015}. While using low-level editing operations for complex 3D shape modeling is powerful, it can be extremely tedious and very challenging for novice users. Fortunately, many tools and systems have been suggested to assist both professional and hobbyist designers during modeling or prototyping \cite{olsen2009sketch,Kazmi2014}. In this chapter we examine different modeling tools and shape representations while focusing on ones that permit easy manipulation of the shape's topology.

\subsection{Implicit Surface Modeling}

\begin{figure}[t!]
	\centering
	\includegraphics[width=0.99\linewidth]{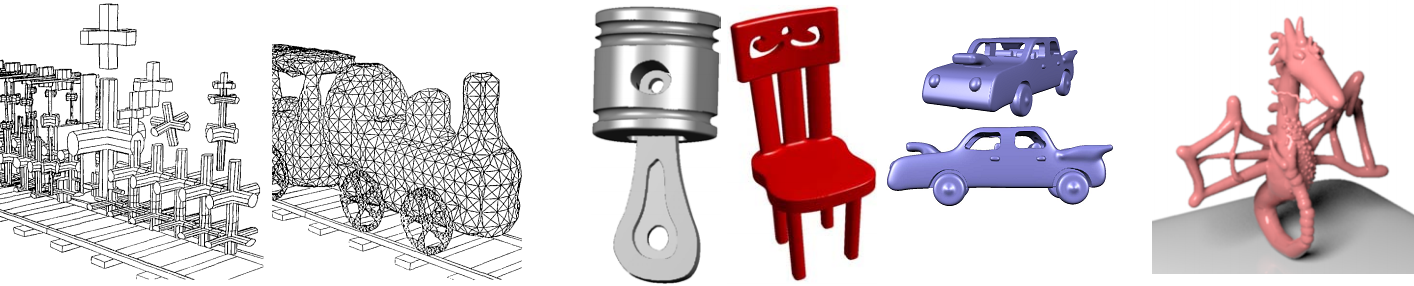}
	\caption{Implicit surface modeling. (left) A train defined by skeletal elements that define a blended surface \protect\cite{Bloomenthal1990}. (middle) The ShapeShope system is a modeling-by-sketching tool that uses implicit surfaces as its shape representation \protect\cite{Schmidt2006}. (left) Later developments on the blending functions allowed for more precise modeling \protect\cite{Bernhardt2010}.  }
	\label{fig:modeling-implicit}
\end{figure}

The technique of Bloomenthal and Wyvill \cite{Bloomenthal1990} allows interactive design of blended \emph{implicit surfaces} defined by skeletal elements (see Figure \ref{fig:modeling-implicit}(left)). The user defines a skeleton from a set of disconnected geometric primitives, such as curves or polygons, and the system automatically generates offset surfaces that are blended together (with some fine control \cite{Bernhardt2010}; Figure \ref{fig:modeling-implicit}(right)). Later work \cite{Schmidt2006} focused on simplifying the creation and editing of these models via sketching of the primitives (Figure \ref{fig:modeling-implicit}(middle)). Complex shapes of almost any topology can be created by constructing a hierarchy of primitives combined with modeling operators as in CSG Trees \cite{Requicha1980}. Another notable example of implicit surface modeling is the F-rep representation described in \cite{Pasko1995}. This technique was later used for morphing shapes of arbitrary topology \cite{Pasko2004}.

\subsection{Interactive Modeling of Surface Meshes}

Polygonal meshes are the most common shape representation used in 3D modeling \cite{PMP2010}. The modeling process often entails the modification of existing polygons by ways of transforming, splitting, or extruding. 

\begin{figure}[t!]
	\centering
	\includegraphics[width=0.99\linewidth]{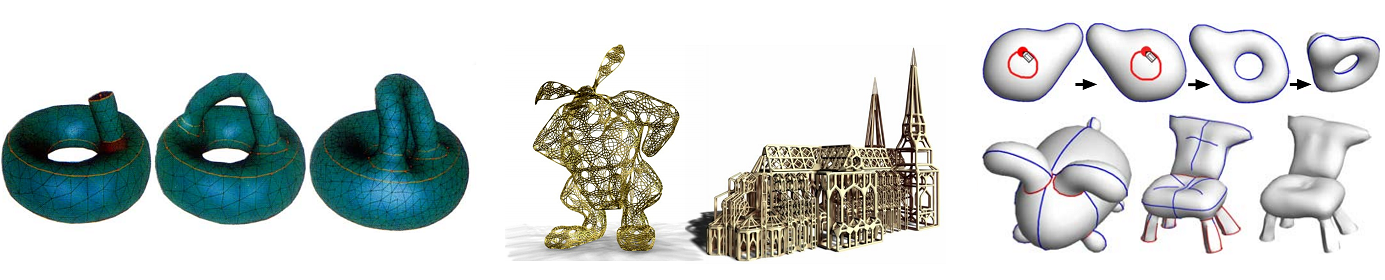}
	\caption{Interactive modeling of surface meshes. (left) A technique for modeling free-form 3D meshes of mutable topology \protect\cite{Welch1994}. (middle) The Topmod3d system is a specialized tool that helps users in modeling highly complex subdivision surfaces \protect\cite{Akleman2008}. (right) The FiberMesh system allows the user to change the topology of the shape with a single stroke. \protect\cite{Nealen2007}.  }
	\label{fig:modeling-surface}
\end{figure}

An early example \cite{Welch1994} of an interactive editing system allowed immediate topological changes for free-form surface design controlled simply by closed curves (Figure \ref{fig:modeling-surface}(left)). The ability to easily modify topology was proposed for subdivision surfaces in \cite{Akleman2000}; the work was later developed into a full modeling system \cite{Akleman2008} that enabled users to create topologically very complex shapes (Figure \ref{fig:modeling-surface}(middle)). Such systems however are still restricted to a set of operations that are not intuitive for novice users to control.

Other more direct and simplified systems that produce polygonal meshes uses sketching \cite{olsen2009sketch,Kazmi2014}. The famous sketch-based modeling system Teddy \cite{Igarashi1999} allows users to easily model free-form 3D surfaces from 2D sketches, however, topology changes were not handled. A later extension described in \cite{Nealen2007} included a tunneling operator that allows users to model handles which modify the topology of the shape (Figure \ref{fig:modeling-surface}(right)). While sketch-based modeling is perhaps the simplest paradigm, it is not ideal for the creation of accurate and topologically complex 3D shapes.

\subsection{Cut-and-paste Tools}

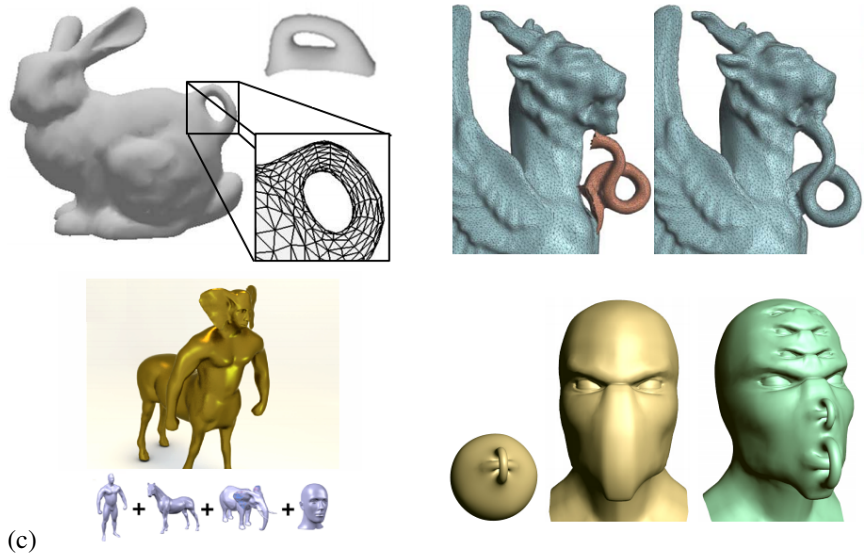
\begin{figure}[t!]
	\centering

	\begin{overpic}[width=0.99\linewidth,tics=10]{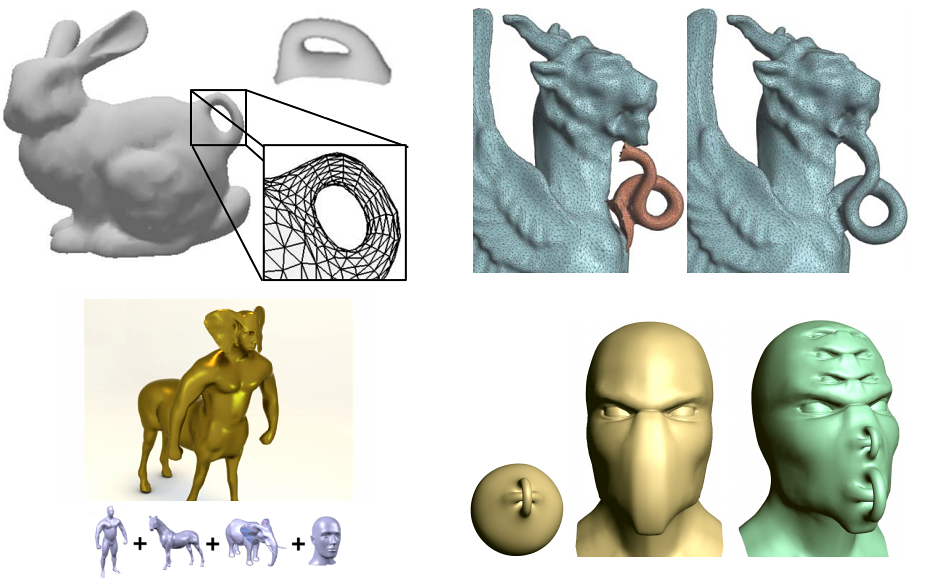}
		\put(0,145){\colorbox{white}{(a)}}
		\put(210,145){\colorbox{white}{(b)}}
		\put(0,0){(c)}
		\put(210,0){(d)}
	\end{overpic}
	
	\caption{Cut-and-paste tools. (a) Non-zero genus part cutting and pasting \protect\cite{Fu2004}. (b) The SnapPaste system is used to attach a tail part into a beard. \protect\cite{Sharf2006}. (c) A combination of four models merged together using the meshmixer system \protect\cite{Schmidt2010}. (d) The GeoBrush system allows the cloning of geometric details including those with complex topology \protect\cite{Takayama2011}.}
	\label{fig:modeling-paste}
\end{figure}

The intuitive cut-and-paste paradigm is ubiquitous in different design applications. Early methods generally focused on connecting 3D meshes by seamlessly gluing, or bridging, boundaries of two disconnected parts using a smooth patch. Later work was focused on finding the best placement with minimal distortion to the pasted parts. The method of Fu et al. \cite{Fu2004} introduced a cut-and-paste technique that allowed the pasting of non-zero genus parts with minimum distortion (Figure \ref{fig:modeling-paste}(a)). The SnapPaste system described in \cite{Sharf2006} focuses on minimizing distortions due to pasting by finding the best alignment between the parts while also considering multiple boundaries (Figure \ref{fig:modeling-paste}(b)).   

A more successful effort is the \emph{meshmixer} system first introduced in \cite{Schmidt2010}. The system provides the user with intuitive tools for shape composition, including part fusion and detail transfer, which enables the rapid creation of complex shapes (Figure \ref{fig:modeling-paste}(c)). Due to the automation of different modeling tasks and its intuitive interface, the system gained wide adoption and was later incorporated into a commercial product \cite{Meshmixer2015}. The pasting of complex surface details that includes handles was also proposed in the \emph{GeoBrush} interactive 3D cloning system \cite{Takayama2011} (see Figure \ref{fig:modeling-paste}(d)).

\subsection{Sculpting}

\begin{figure}[t!]
	\centering
	\includegraphics[width=0.99\linewidth]{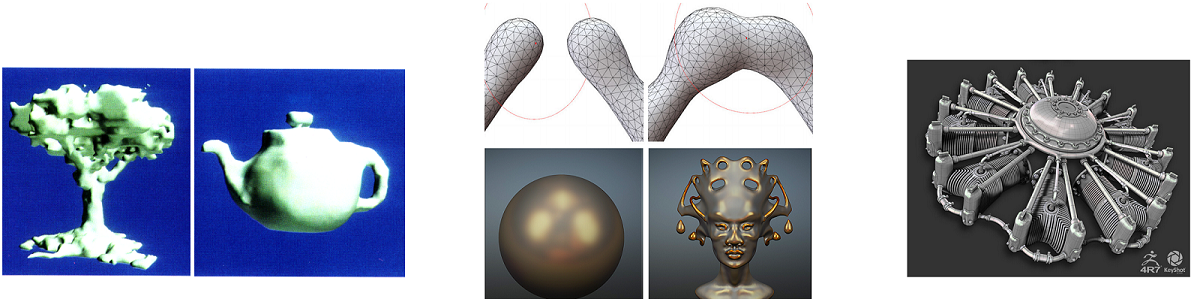}
	\caption{Sculpting. (left) An early system for sculpting 3D shapes by manipulating voxel arrays \protect\cite{Galyean1991}. (middle) Later sculpting systems use mesh-based surface representations while naturally handling topological changes \protect\cite{Stanculescu2011}. (right) The latest version of ZBrush incorporates many tools that allow for extremely complicated and detailed models \protect\cite{ZBrush2015}.}
	\label{fig:modeling-sculpting}
\end{figure}

Digital sculpting has become a film and game industry standard for shape creation and 3D modeling \cite{ZBrush2015,Mudbox2014}. A typical 3D sculpting tool provides the user with a set of shape primitives, such as spheres or cylinders, and a set of controllable sculpting tools that include pushing or pulling on the shape to insert and modify surface details \cite{Parent1977,Galyean1991} (Figure \ref{fig:modeling-sculpting}(left)). 

Several representations have been proposed for digital sculpting. In Museth et al. \cite{Museth2002}, level set surfaces were used in a free-form sculpting system with many tool that easily allows change in topology. The point cloud representation in \cite{Pauly2003} allows for a more efficient shape modeling system with flexible controls over self-intersections behavior. The volume preserving sculpting of \cite{Angelidis2006} allows unlimited stretching of meshes, however, it only avoids self-intersections with no support for topology change. The \emph{B-Mesh} system introduced in \cite{Ji2010} simplifies the process of creating base polygonal meshes for sculpting by using a skeletal shape balls representation that can be easily articulated. An efficient mesh-based sculpting system presented in \cite{Stanculescu2011} handles arbitrary changes in topology through an adaptive sampling of the surface (Figure \ref{fig:modeling-sculpting}(middle)). The most recent version of the commercial software \emph{ZBrush} incorporates almost all major advancement in the area of digital sculpting (Figure \ref{fig:modeling-sculpting}(right)). 

Other non-traditional sculpting tools include the \emph{smart-clay} system presented in \cite{Milliez2013} which tried to extend the sculpting paradigm to the area of structured shapes modeling which supports shapes of arbitrary topology. The work of \cite{Bernstein2013} is the first method that allows sculpting with topological changes on any mesh surfaces including ones that are non-manifold and with self-intersection. The work of \cite{Baerentzen14} defines a dual representation where sculpting operations include both surface and skeletal modifying modeling operations.

\subsection{Mesh Repair and Topology Modification}

While modeling generally involves evolving shapes into more complex forms, it is sometimes desirable to appropriately simplify shape complexity as in mesh repair applications \cite{Attene2013}. Surface reconstruction from noisy medical imaging often produce topological errors. The work of Wood et al. \cite{Wood2004} tries to remove excess topology by selectively removing less prominent handles in an isosurface. A later method in \cite{Zhou2007} focuses on repairing topological errors using operations on the shape's skeleton. The \emph{Morfit} method in \cite{Yin2014} is an interactive system where the user can assist in the reconstruction process by modifying skeleton branches and profile curves.

\subsection{Topologically Complex Geometric Detail}

The work of Peng et al. \cite{Peng2004} uses aligned volumetric textures in order to model and render topologically complex geometric details. However, the main use of pixel-based methods is the rendering of details which are not suitable for further editing or physical fabrication. The mesh-based quilting method from Zhou et al. \cite{Zhou2006} results in well stitched 3D geometry elements. While such approaches can model topologically complex surfaces, they are mainly intended for attaching surface details and not for general purpose modeling.

\subsection{Shape Parts Graphs}

Tools that permit the modification of the shape's topology provide low-level operations which are not always intuitive for novice users. The most basic modeling paradigm is that of part assembly where users combine many existing parts to construct topologically complex 3D objects. A recent trend in shape modeling is the use of part patterns and structural shape graphs \cite{Mitra13}. These graphs often encode how parts connect with each other along with different structural relations. The modeling process can then be simplified by modifying graph attributes rather than geometric elements. 

The inverse procedural modeling work of Bokeloh et al. \cite{Bokeloh2010} allows the creation of complicated forms of an exemplar shape by analyzing repeated patterns and identifying potential docking sites where parts connect. The symmetry hierarchy work of Wang et al. \cite{Wang11} proposed a structural organization of parts where different parts are grouped by symmetry or by assembly. Later methods also used symmetry analysis to identify functional sub-structures allowing the modeling of functionally plausible shapes by simply shuffling a group of parts \cite{Zheng2013}. Part relation graphs were also used in Zheng et al. \cite{Zheng14} for shape correspondence of very different models, thus, demonstrating the potential of rich structural shape graphs in 3D modeling.

	
	\bibliographystyle{acmsiggraph}
	\bibliography{report}

\begin{thebibliography}{\protect\citename{B{\ae}rentzen et~al\mbox{.} }2014}

\bibitem[\protect\citename{Aiger et~al\mbox{.} }2008]{Aiger2008}
{\sc Aiger, D., Mitra, N.~J., and Cohen-Or, D.}
\newblock 2008.
\newblock 4pointss congruent sets for robust pairwise surface registration.
\newblock {\em ACM Trans. Graph. 27}, 3 (Aug.), 85:1--85:10.

\bibitem[\protect\citename{Akleman et~al\mbox{.} }2000]{Akleman2000}
{\sc Akleman, E., Chen, J., and Srinivasan, V.}
\newblock 2000.
\newblock A new paradigm for changing topology during subdivision modeling.
\newblock In {\em Computer Graphics and Applications, 2000. Proceedings. The
  Eighth Pacific Conference on}, 192--201.

\bibitem[\protect\citename{Akleman et~al\mbox{.} }2008]{Akleman2008}
{\sc Akleman, E., Srinivasan, V., Chen, J., Morris, D., and Tett, S.}
\newblock 2008.
\newblock Topmod3d: An interactive topological mesh modeler.
\newblock In {\em Proc. of CGI}, 10--18.

\bibitem[\protect\citename{Alexa }2002]{Alexa02}
{\sc Alexa, M.}
\newblock 2002.
\newblock Recent advances in mesh morphing.
\newblock {\em Computer Graphics Forum 21}, 2, 173--198.

\bibitem[\protect\citename{Angelidis et~al\mbox{.} }2006]{Angelidis2006}
{\sc Angelidis, A., Cani, M.-P., Wyvill, G., and King, S.}
\newblock 2006.
\newblock Swirling-sweepers: Constant-volume modeling.
\newblock {\em Graphical Models 68}, 4, 324 -- 332.
\newblock \{PG2004\}.

\bibitem[\protect\citename{Athanasiadis et~al\mbox{.} }2012]{Athanasiadis2012}
{\sc Athanasiadis, T., Fudos, I., Nikou, C., and Stamati, V.}
\newblock 2012.
\newblock Feature-based 3d morphing based on geometrically constrained
  spherical parameterization.
\newblock {\em Computer Aided Geometric Design 29}, 1, 2 -- 17.
\newblock Geometric Constraints and Reasoning.

\bibitem[\protect\citename{Attene et~al\mbox{.} }2013]{Attene2013}
{\sc Attene, M., Campen, M., and Kobbelt, L.}
\newblock 2013.
\newblock Polygon mesh repairing: An application perspective.
\newblock {\em ACM Comput. Surv. 45}, 2 (Mar.), 15:1--15:33.

\bibitem[\protect\citename{Au et~al\mbox{.} }2008]{Au08}
{\sc Au, O. K.-C., Tai, C.-L., Chu, H.-K., and Cohen-Or, D.}
\newblock 2008.
\newblock Skeleton extraction by mesh contraction.
\newblock {\em ACM Trans. on Graphics 27}, 3.

\bibitem[\protect\citename{Au et~al\mbox{.} }2010]{Au10}
{\sc Au, O. K.-C., Cohen-Or, D., Tai, C.-L., Fu, H., and Zheng, Y.}
\newblock 2010.
\newblock Electors voting for fast automatic shape correspondence.
\newblock {\em Computer Graphics Forum (Proc. EUROGRAPHICS) 29}, 2.

\bibitem[\protect\citename{B{\ae}rentzen et~al\mbox{.} }2014]{Baerentzen14}
{\sc B{\ae}rentzen, J.~A., Abdrashitov, R., and Singh, K.}
\newblock 2014.
\newblock Interactive shape modeling using a skeleton-mesh co-representation.
\newblock {\em ACM Transactions on Graphics (proceedings of ACM SIGGRAPH) 33},
  4.

\bibitem[\protect\citename{Baumgart }1974]{Baumgart74}
{\sc Baumgart, B.~G.}
\newblock 1974.
\newblock {\em Geometric Modeling for Computer Vision.}
\newblock PhD thesis, Stanford, CA, USA.
\newblock AAI7506806.

\bibitem[\protect\citename{Beier and Neely }1992]{Beier1992}
{\sc Beier, T., and Neely, S.}
\newblock 1992.
\newblock Feature-based image metamorphosis.
\newblock In {\em Computer Graphics (SIGGRAPH '92 Proceedings)}, vol.~26,
  35--42.

\bibitem[\protect\citename{Bendsoe }2003]{Bendsoe2003}
{\sc Bendsoe, M.}
\newblock 2003.
\newblock {\em Topology Optimization: {Theory,Methods} and Applications}.
\newblock Springer-Verlag, Berlin; Heidelberg; New York.

\bibitem[\protect\citename{Bernhardt et~al\mbox{.} }2010]{Bernhardt2010}
{\sc Bernhardt, A., Barthe, L., Cani, M.-P., and Wyvill, B.}
\newblock 2010.
\newblock Implicit blending revisited.
\newblock In {\em Computer Graphics Forum}, vol.~29, Wiley Online Library,
  367--375.

\bibitem[\protect\citename{Bernstein and Wojtan }2013]{Bernstein2013}
{\sc Bernstein, G.~L., and Wojtan, C.}
\newblock 2013.
\newblock Putting holes in holey geometry: Topology change for arbitrary
  surfaces.
\newblock {\em ACM Trans. Graph. 32}, 4 (July), 34:1--34:12.

\bibitem[\protect\citename{Besl and Mckay }1992]{Besl1992}
{\sc Besl, P.~J., and Mckay, N.~D.}
\newblock 1992.
\newblock A method for registration of 3-d shapes.
\newblock {\em IEEE PAMI 14}, 2, 239--256.

\bibitem[\protect\citename{Biasotti et~al\mbox{.} }2006]{Biasotti06}
{\sc Biasotti, S., Marini, S., Spagnuolo, M., and Falcidieno, B.}
\newblock 2006.
\newblock Sub-part correspondence by structural descriptors of {3D} shapes.
\newblock {\em Computer-Aided Design 38}, 9, 1002--1019.

\bibitem[\protect\citename{Biasotti et~al\mbox{.} }2008]{Biasotti2008}
{\sc Biasotti, S., De~Floriani, L., Falcidieno, B., Frosini, P., Giorgi, D.,
  Landi, C., Papaleo, L., and Spagnuolo, M.}
\newblock 2008.
\newblock Describing shapes by geometrical-topological properties of real
  functions.
\newblock {\em ACM Comput. Surv. 40}, 4 (Oct.), 12:1--12:87.

\bibitem[\protect\citename{Blanding et~al\mbox{.} }2000]{Blanding2000}
{\sc Blanding, R.~L., Turkiyyah, G.~M., Storti, D.~W., and Ganter, M.~A.}
\newblock 2000.
\newblock Skeleton-based three-dimensional geometric morphing.
\newblock {\em Computational Geometry 15}, 1–3, 129 -- 148.

\bibitem[\protect\citename{Bloomenthal and Wyvill }1990]{Bloomenthal1990}
{\sc Bloomenthal, J., and Wyvill, B.}
\newblock 1990.
\newblock Interactive techniques for implicit modeling.
\newblock {\em SIGGRAPH Comput. Graph. 24}, 2 (Feb.), 109--116.

\bibitem[\protect\citename{Blum }1967]{Blum1967}
{\sc Blum, H.}
\newblock 1967.
\newblock {A Transformation for Extracting New Descriptors of Shape}.
\newblock {\em Models for the Perception of Speech and Visual Form\/},
  362--380.

\bibitem[\protect\citename{Bokeloh et~al\mbox{.} }2010]{Bokeloh2010}
{\sc Bokeloh, M., Wand, M., and Seidel, H.-P.}
\newblock 2010.
\newblock A connection between partial symmetry and inverse procedural
  modeling.
\newblock {\em ACM Trans. Graph. 29}, 4 (July), 104:1--104:10.

\bibitem[\protect\citename{Botsch et~al\mbox{.} }2010]{PMP2010}
{\sc Botsch, M., Kobbelt, L., Pauly, M., Alliez, P., and Levy, B.}
\newblock 2010.
\newblock {\em Polygon Mesh Processing}.
\newblock AK Peters.

\bibitem[\protect\citename{Breen and Whitaker }2001]{Breen2001}
{\sc Breen, D., and Whitaker, R.}
\newblock 2001.
\newblock A level-set approach for the metamorphosis of solid models.
\newblock {\em Visualization and Computer Graphics, IEEE Transactions on 7}, 2
  (Apr), 173--192.

\bibitem[\protect\citename{Bronstein et~al\mbox{.} }2008]{Bronstein08}
{\sc Bronstein, A.~M., Bronstein, M.~M., Bruckstein, A.~M., and Kimmel, R.}
\newblock 2008.
\newblock Partial similarity of objects, or how to compare a centaur to a
  horse.
\newblock {\em Int. J. of Computer Vision\/}.

\bibitem[\protect\citename{Carbonell and Goldstein }1998]{Carbonell1998}
{\sc Carbonell, J., and Goldstein, J.}
\newblock 1998.
\newblock The use of mmr, diversity-based reranking for reordering documents
  and producing summaries.
\newblock In {\em Proceedings of the 21st Annual International ACM SIGIR
  Conference on Research and Development in Information Retrieval}, ACM, New
  York, NY, USA, SIGIR '98, 335--336.

\bibitem[\protect\citename{Chang et~al\mbox{.} }2011]{Chang2011}
{\sc Chang, W., Li, H., Mitra, N.~J., Pauly, M., Rusinkiewicz, S., and Wand,
  M.}
\newblock 2011.
\newblock Computing correspondences in geometric data sets.
\newblock In {\em Eurographics 2011: Tutorial Notes}.

\bibitem[\protect\citename{Chaudhuri and Koltun }2010]{Chaudhuri2010}
{\sc Chaudhuri, S., and Koltun, V.}
\newblock 2010.
\newblock Data-driven suggestions for creativity support in 3d modeling.
\newblock {\em ACM Trans. Graph. 29}, 6 (Dec.), 183:1--183:10.

\bibitem[\protect\citename{Chaudhuri et~al\mbox{.} }2011]{Chaudhuri2011}
{\sc Chaudhuri, S., Kalogerakis, E., Guibas, L., and Koltun, V.}
\newblock 2011.
\newblock Probabilistic reasoning for assembly-based 3d modeling.
\newblock In {\em ACM SIGGRAPH 2011 Papers}, ACM, New York, NY, USA, SIGGRAPH
  '11, 35:1--35:10.

\bibitem[\protect\citename{Chen and Medioni }1991]{Chen1991}
{\sc Chen, Y., and Medioni, G.}
\newblock 1991.
\newblock Object modeling by registration of multiple range images.
\newblock In {\em Robotics and Automation, 1991. Proceedings., 1991 IEEE
  International Conference on}, 2724--2729 vol.3.

\bibitem[\protect\citename{Chen and Parent }1989]{Chen1989}
{\sc Chen, S., and Parent, R.}
\newblock 1989.
\newblock Shape averaging and its applications to industrial design.
\newblock {\em Computer Graphics and Applications, IEEE 9}, 1 (Jan), 47--54.

\bibitem[\protect\citename{Chen et~al\mbox{.} }2003]{Chen2003}
{\sc Chen, L.-L., Wang, G.~F., Hsiao, K.-A., and Liang, J.}
\newblock 2003.
\newblock Affective product shapes through image morphing.
\newblock In {\em Proceedings of the 2003 International Conference on Designing
  Pleasurable Products and Interfaces}, ACM, New York, NY, USA, DPPI '03,
  11--16.

\bibitem[\protect\citename{Cohen-Or et~al\mbox{.} }1996]{Cohen1996}
{\sc Cohen-Or, D., Levin, D., and Solomovici, A.}
\newblock 1996.
\newblock Contour blending using warp-guided distance field interpolation.
\newblock In {\em Visualization '96. Proceedings.}, 165--172.

\bibitem[\protect\citename{Cohen-Or et~al\mbox{.} }1998]{Cohen1998}
{\sc Cohen-Or, D., Solomovic, A., and Levin, D.}
\newblock 1998.
\newblock Three-dimensional distance field metamorphosis.
\newblock {\em ACM Trans. Graph. 17}, 2 (Apr.), 116--141.

\bibitem[\protect\citename{DAZ-Studio }2014]{DAZ2014}
{\sc DAZ-Studio}, 2014.
\newblock DAZ 3D, \texttt{http://www.daz3d.com/technology/}.

\bibitem[\protect\citename{DeCarlo and Gallier }1996]{DeCarlo96}
{\sc DeCarlo, D., and Gallier, J.}
\newblock 1996.
\newblock Topological evolution of surfaces.
\newblock In {\em Proceedings of GI'96}, Canadian Human-Computer Communications
  Society, 194--203.

\bibitem[\protect\citename{Elad and Kimmel }2003]{Elad03}
{\sc Elad, A., and Kimmel, R.}
\newblock 2003.
\newblock On bending invariant signatures for surfaces.
\newblock {\em IEEE PAMI 25}, 10, 1285--1295.

\bibitem[\protect\citename{Fu et~al\mbox{.} }2004]{Fu2004}
{\sc Fu, H., Tai, C.-L., and Zhang, H.}
\newblock 2004.
\newblock Topology-free cut-and-paste editing over meshes.
\newblock In {\em Geometric Modeling and Processing, 2004. Proceedings},
  173--182.

\bibitem[\protect\citename{Funkhouser et~al\mbox{.} }2004]{Funkhouser04}
{\sc Funkhouser, T., Kazhdan, M., Shilane, P., Min, P., Kiefer, W., Tal, A.,
  Rusinkiewicz, S., and Dobkin, D.}
\newblock 2004.
\newblock Modeling by example.
\newblock {\em ACM Trans. on Graphics 23}, 3, 652--663.

\bibitem[\protect\citename{Galyean and Hughes }1991]{Galyean1991}
{\sc Galyean, T.~A., and Hughes, J.~F.}
\newblock 1991.
\newblock Sculpting: An interactive volumetric modeling technique.
\newblock {\em SIGGRAPH Comput. Graph. 25}, 4 (July), 267--274.

\bibitem[\protect\citename{Golovinskiy and Funkhouser }2008]{Golovinskiy2008}
{\sc Golovinskiy, A., and Funkhouser, T.}
\newblock 2008.
\newblock Randomized cuts for 3d mesh analysis.
\newblock {\em ACM Trans. Graph. 27}, 5 (Dec.), 145:1--145:12.

\bibitem[\protect\citename{Golovinskiy and Funkhouser }2009]{Golovinskiy09}
{\sc Golovinskiy, A., and Funkhouser, T.}
\newblock 2009.
\newblock Consistent segmentation of {3D} models.
\newblock {\em Computers \& Graphics (Proc. SMI) 33}, 3, 262--269.

\bibitem[\protect\citename{Gomes et~al\mbox{.} }1998]{Gomes1998}
{\sc Gomes, J., Darsa, L., Costa, B., and Velho, L.}
\newblock 1998.
\newblock {\em Warping and Morphing of Graphical Objects}.
\newblock Morgan Kaufmann Publishers Inc., San Francisco, CA, USA.

\bibitem[\protect\citename{He et~al\mbox{.} }1994]{He1994}
{\sc He, T., Wang, S., and Kaufman, A.}
\newblock 1994.
\newblock Wavelet-based volume morphing.
\newblock In {\em Visualization, 1994., Visualization '94, Proceedings., IEEE
  Conference on}, 85--92, CP8.

\bibitem[\protect\citename{Hilaga et~al\mbox{.} }2001]{Hilaga01}
{\sc Hilaga, M., Shinagawa, Y., Kohmura, T., and Kunii, T.~L.}
\newblock 2001.
\newblock Topology matching for fully automatic similarity estimation of {3D}
  shapes.
\newblock In {\em Proc. SIGGRAPH}, 203--212.

\bibitem[\protect\citename{Houston et~al\mbox{.} }2006]{Houston2006}
{\sc Houston, B., Nielsen, M.~B., Batty, C., Nilsson, O., and Museth, K.}
\newblock 2006.
\newblock Hierarchical rle level set: A compact and versatile deformable
  surface representation.
\newblock {\em ACM Trans. Graph. 25}, 1, 151--175.

\bibitem[\protect\citename{Huang et~al\mbox{.} }2008]{Huang08}
{\sc Huang, Q.-X., Adams, B., Wicke, M., and Guibas, L.~J.}
\newblock 2008.
\newblock Non-rigid registration under isometric deformations.
\newblock {\em Computer Graphics Forum (Proc. SGP) 27}, 5, 1449--1457.

\bibitem[\protect\citename{Huang et~al\mbox{.} }2014]{Huang2014}
{\sc Huang, Q., Wang, F., and Guibas, L.}
\newblock 2014.
\newblock Functional map networks for analyzing and exploring large shape
  collections.
\newblock {\em ACM Trans. Graph. 33}, 4 (July), 36:1--36:11.

\bibitem[\protect\citename{Hughes }1992]{Hughes1992}
{\sc Hughes, J.~F.}
\newblock 1992.
\newblock Scheduled fourier volume morphing.
\newblock {\em SIGGRAPH Comput. Graph. 26}, 2 (July), 43--46.

\bibitem[\protect\citename{Hui and Li }1998]{Hui1998}
{\sc Hui, K., and Li, Y.}
\newblock 1998.
\newblock A feature-based shape blending technique for industrial design.
\newblock {\em Computer-Aided Design 30}, 10, 823 -- 834.

\bibitem[\protect\citename{Igarashi et~al\mbox{.} }1999]{Igarashi1999}
{\sc Igarashi, T., Matsuoka, S., and Tanaka, H.}
\newblock 1999.
\newblock Teddy: A sketching interface for 3d freeform design.
\newblock In {\em Proceedings of the 26th Annual Conference on Computer
  Graphics and Interactive Techniques}, ACM Press/Addison-Wesley Publishing
  Co., New York, NY, USA, SIGGRAPH '99, 409--416.

\bibitem[\protect\citename{Jain and Zhang }2006]{jain2006robust}
{\sc Jain, V., and Zhang, H.}
\newblock 2006.
\newblock Robust 3d shape correspondence in the spectral domain.
\newblock In {\em Shape Modeling and Applications, 2006. SMI 2006. IEEE
  International Conference on}, IEEE, 19--19.

\bibitem[\protect\citename{Jain et~al\mbox{.} }2012]{Jain2012}
{\sc Jain, A., Thormhlen, T., Ritschel, T., and Seidel, H.-P.}
\newblock 2012.
\newblock Exploring shape variations by 3d-model decomposition and part-based
  recombination.
\newblock {\em Comp. Graph. Forum 31}, 2pt3 (May), 631--640.

\bibitem[\protect\citename{Ji et~al\mbox{.} }2010]{Ji2010}
{\sc Ji, Z., Liu, L., and Wang, Y.}
\newblock 2010.
\newblock B-mesh: A modeling system for base meshes of 3d articulated shapes.
\newblock In {\em Computer Graphics Forum}, vol.~29, Wiley Online Library,
  2169--2177.

\bibitem[\protect\citename{Kalogerakis et~al\mbox{.} }2010]{Kalogerakis10}
{\sc Kalogerakis, E., Hertzmann, A., and Singh, K.}
\newblock 2010.
\newblock Learning {3D} mesh segmentation and labeling.
\newblock {\em ACM Trans. on Graphics 29}, 3.

\bibitem[\protect\citename{Kalogerakis et~al\mbox{.} }2012]{Kalogerakis2012}
{\sc Kalogerakis, E., Chaudhuri, S., Koller, D., and Koltun, V.}
\newblock 2012.
\newblock A probabilistic model for component-based shape synthesis.
\newblock {\em ACM Trans. Graph. 31}, 4 (July), 55:1--55:11.

\bibitem[\protect\citename{Kanai et~al\mbox{.} }1997]{Kanai97}
{\sc Kanai, T., Suzuki, H., and Kimura, F.}
\newblock 1997.
\newblock 3d geometric metamorphosis based on harmonic map.
\newblock In {\em Computer Graphics and Applications, 1997. Proceedings., The
  Fifth Pacific Conference on}, 97--104.

\bibitem[\protect\citename{Kanai et~al\mbox{.} }2000]{Kanai2000}
{\sc Kanai, T., Suzuki, H., and Kimura, F.}
\newblock 2000.
\newblock Metamorphosis of arbitrary triangular meshes.
\newblock {\em Computer Graphics and Applications, IEEE 20}, 2 (Mar), 62--75.

\bibitem[\protect\citename{Kang and Lee }2010]{Kang2010}
{\sc Kang, J.~Y., and Lee, B.}
\newblock 2010.
\newblock Mesh-based morphing method for rapid hull form generation.
\newblock {\em Computer-Aided Design 42}, 11, 970 -- 976.
\newblock Computer aided ship design: Some recent results and steps ahead in
  theory, methodology and practice Dedicated to Professor Horst Nowacki on the
  occasion of his 75th birthday.

\bibitem[\protect\citename{Kazmi et~al\mbox{.} }2014]{Kazmi2014}
{\sc Kazmi, I., You, L., and Zhang, J.~J.}
\newblock 2014.
\newblock A survey of sketch based modeling systems.
\newblock In {\em Computer Graphics, Imaging and Visualization (CGIV), 2014
  11th International Conference on}, 27--36.

\bibitem[\protect\citename{Kent et~al\mbox{.} }1992]{Kent1992}
{\sc Kent, J.~R., Carlson, W.~E., and Parent, R.~E.}
\newblock 1992.
\newblock Shape transformation for polyhedral objects.
\newblock In {\em Proceedings of the 19th Annual Conference on Computer
  Graphics and Interactive Techniques}, ACM, New York, NY, USA, SIGGRAPH '92,
  47--54.

\bibitem[\protect\citename{Kim et~al\mbox{.} }2011]{Kim11}
{\sc Kim, V.~G., Lipman, Y., and Funkhouser, T.}
\newblock 2011.
\newblock Blended intrinsic maps.
\newblock {\em ACM Trans. on Graphics 30}, 4, 79:1--79:12.

\bibitem[\protect\citename{Kim et~al\mbox{.} }2012]{Kim2012}
{\sc Kim, V.~G., Li, W., Mitra, N.~J., DiVerdi, S., and Funkhouser, T.}
\newblock 2012.
\newblock Exploring collections of 3d models using fuzzy correspondences.
\newblock {\em ACM Transactions on Graphics (TOG) 31}, 4, 54.

\bibitem[\protect\citename{Kim et~al\mbox{.} }2013]{Kim2013}
{\sc Kim, V.~G., Li, W., Mitra, N.~J., Chaudhuri, S., DiVerdi, S., and
  Funkhouser, T.}
\newblock 2013.
\newblock Learning part-based templates from large collections of 3d shapes.
\newblock {\em ACM Transactions on Graphics (TOG) 32}, 4, 70.

\bibitem[\protect\citename{Kraevoy and Sheffer }2004]{Kraevoy2004}
{\sc Kraevoy, V., and Sheffer, A.}
\newblock 2004.
\newblock Cross-parameterization and compatible remeshing of 3d models.
\newblock {\em ACM Trans. Graph. 23}, 3 (Aug.), 861--869.

\bibitem[\protect\citename{Kreavoy et~al\mbox{.} }2007]{Kreavoy2007}
{\sc Kreavoy, V., Julius, D., and Sheffer, A.}
\newblock 2007.
\newblock Model composition from interchangeable components.
\newblock In {\em Proceedings of the 15th Pacific Conference on Computer
  Graphics and Applications}, IEEE Computer Society, Washington, DC, USA, PG
  '07, 129--138.

\bibitem[\protect\citename{Lazarus and Verroust }1998]{Lazarus98}
{\sc Lazarus, F., and Verroust, A.}
\newblock 1998.
\newblock Three-dimensional metamorphosis: a survey.
\newblock {\em The Visual Computer 14}, 8-9, 373--389.

\bibitem[\protect\citename{Lee et~al\mbox{.} }1999]{Lee1999}
{\sc Lee, A. W.~F., Dobkin, D., Sweldens, W., and Schr\"{o}der, P.}
\newblock 1999.
\newblock Multiresolution mesh morphing.
\newblock In {\em Proceedings of the 26th Annual Conference on Computer
  Graphics and Interactive Techniques}, ACM Press/Addison-Wesley Publishing
  Co., New York, NY, USA, SIGGRAPH '99, 343--350.

\bibitem[\protect\citename{Lee et~al\mbox{.} }2006]{Lee2006}
{\sc Lee, T.-Y., Yao, C.-Y., Chu, H.-K., Tai, M.-J., and Chen, C.-C.}
\newblock 2006.
\newblock Generating genus-n-to-m mesh morphing using spherical
  parameterization.
\newblock {\em Computer Animation and Virtual Worlds 17}, 3-4, 433--443.

\bibitem[\protect\citename{Levin }1986]{Levin1986}
{\sc Levin, D.}
\newblock 1986.
\newblock Multidimensional reconstruction by set-valued approximations.
\newblock {\em IMA Journal of Numerical Analysis 6}, 2, 173--184.

\bibitem[\protect\citename{Li and Guskov }2005]{Li2005multiscale}
{\sc Li, X., and Guskov, I.}
\newblock 2005.
\newblock Multiscale features for approximate alignment of point-based
  surfaces.
\newblock In {\em Symposium on geometry processing}, vol.~2, Citeseer.

\bibitem[\protect\citename{Li et~al\mbox{.} }2011]{Li2011}
{\sc Li, B., Liu, X., Liu, Y., Hu, P., Liu, M., and Wang, C.}
\newblock 2011.
\newblock Image-driven panel design via feature-preserving mesh deformation.
\newblock In {\em Computer and Computing Technologies in Agriculture IV},
  D.~Li, Y.~Liu, and Y.~Chen, Eds., vol.~345 of {\em IFIP Advances in
  Information and Communication Technology}. Springer Berlin Heidelberg,
  30--40.

\bibitem[\protect\citename{Liao et~al\mbox{.} }2014a]{Liao2014b}
{\sc Liao, J., Lima, R.~S., Nehab, D., Hoppe, H., and Sander, P.~V.}
\newblock 2014.
\newblock Semi-automated video morphing.
\newblock {\em Computer Graphics Forum 33}, 4, 51--60.

\bibitem[\protect\citename{Liao et~al\mbox{.} }2014b]{Liao2014}
{\sc Liao, J., Lima, R.~S., Nehab, D., Hoppe, H., Sander, P.~V., and Yu, J.}
\newblock 2014.
\newblock Automating image morphing using structural similarity on a halfway
  domain.
\newblock {\em ACM Trans. Graph. 33}, 5 (Sept.), 168:1--168:12.

\bibitem[\protect\citename{Lien and Amato }2008]{Lien2008}
{\sc Lien, J.-M., and Amato, N.~M.}
\newblock 2008.
\newblock Approximate convex decomposition of polyhedra and its applications.
\newblock {\em Comput. Aided Geom. Des. 25}, 7 (Oct.), 503--522.

\bibitem[\protect\citename{Lipman and Funkhouser }2009]{Lipman09}
{\sc Lipman, Y., and Funkhouser, T.}
\newblock 2009.
\newblock M\"{o}bius voting for surface correspondence.
\newblock {\em ACM Trans. on Graphics 28}, 3.

\bibitem[\protect\citename{Liu et~al\mbox{.} }2004]{Liu04}
{\sc Liu, L., Wang, G., Zhang, B., Guo, B., and Shum, H.-Y.}
\newblock 2004.
\newblock Perceptually based approach for planar shape morphing.
\newblock In {\em Proc. of Pacific Graphics}, 111--120.

\bibitem[\protect\citename{Liu et~al\mbox{.} }2005]{Liu2005}
{\sc Liu, L.-G., Zhang, B., Guo, B.-N., and Shum, H.-Y.}
\newblock 2005.
\newblock Polygonal shape blending with topological evolutions.
\newblock {\em Journal of Computer Science and Technology 20}, 1, 77--89.

\bibitem[\protect\citename{Maya }2015]{Maya2015}
{\sc Maya}, 2015.
\newblock Autodesk, \texttt{http://www.autodesk.com/products/maya}.

\bibitem[\protect\citename{Meshmixer }2015]{Meshmixer2015}
{\sc Meshmixer}, 2015.
\newblock Autodesk, \texttt{http://www.123dapp.com/meshmixer}.

\bibitem[\protect\citename{Michikawa et~al\mbox{.} }2001]{Michikawa2001}
{\sc Michikawa, T., Kanai, T., Fujita, M., and Chiyokura, H.}
\newblock 2001.
\newblock Multiresolution interpolation meshes.
\newblock In {\em Computer Graphics and Applications, 2001. Proceedings. Ninth
  Pacific Conference on}, 60--69.

\bibitem[\protect\citename{Milliez et~al\mbox{.} }2013]{Milliez2013}
{\sc Milliez, A., Wand, M., Cani, M.-P., and Seidel, H.-P.}
\newblock 2013.
\newblock Mutable elastic models for sculpting structured shapes.
\newblock {\em Comput. Graph. Forum 32}, 2, 21--30.

\bibitem[\protect\citename{Mitra et~al\mbox{.} }2013]{Mitra13}
{\sc Mitra, N., Wand, M., Zhang, H.~R., Cohen-Or, D., Kim, V., and Huang,
  Q.-X.}
\newblock 2013.
\newblock Structure-aware shape processing.
\newblock In {\em SIGGRAPH Asia 2013 Courses}, 1:1--1:20.

\bibitem[\protect\citename{Mudbox }2014]{Mudbox2014}
{\sc Mudbox}, 2014.
\newblock Autodesk, \texttt{http://www.autodesk.com/products/mudbox}.

\bibitem[\protect\citename{Museth et~al\mbox{.} }2002]{Museth2002}
{\sc Museth, K., Breen, D.~E., Whitaker, R.~T., and Barr, A.~H.}
\newblock 2002.
\newblock Level set surface editing operators.
\newblock {\em ACM Trans. Graph. 21}, 3 (July), 330--338.

\bibitem[\protect\citename{Museth }2013]{Museth2013}
{\sc Museth, K.}
\newblock 2013.
\newblock Vdb: High-resolution sparse volumes with dynamic topology.
\newblock {\em ACM Trans. Graph. 32}, 3 (July), 27:1--27:22.

\bibitem[\protect\citename{Nashvili et~al\mbox{.} }2005]{Nashvili2005}
{\sc Nashvili, M., Olhofer, M., and Sendhoff, B.}
\newblock 2005.
\newblock Morphing methods in evolutionary design optimization.
\newblock In {\em Proceedings of the 7th Annual Conference on Genetic and
  Evolutionary Computation}, ACM, New York, NY, USA, GECCO '05, 897--904.

\bibitem[\protect\citename{Nealen et~al\mbox{.} }2007]{Nealen2007}
{\sc Nealen, A., Igarashi, T., Sorkine, O., and Alexa, M.}
\newblock 2007.
\newblock Fibermesh: designing freeform surfaces with 3d curves.
\newblock In {\em ACM Transactions on Graphics (TOG)}, vol.~26, ACM, 41.

\bibitem[\protect\citename{Norman }2013]{Norman2013}
{\sc Norman, D.~A.}
\newblock 2013.
\newblock {\em The Design of Everyday Things}, revised and expanded edition~ed.
\newblock Nov.

\bibitem[\protect\citename{Olsen et~al\mbox{.} }2009]{olsen2009sketch}
{\sc Olsen, L., Samavati, F.~F., Sousa, M.~C., and Jorge, J.~A.}
\newblock 2009.
\newblock Sketch-based modeling: A survey.
\newblock {\em Computers \& Graphics 33}, 1, 85--103.

\bibitem[\protect\citename{Ovsjanikov et~al\mbox{.} }2011]{Ovsjanikov11}
{\sc Ovsjanikov, M., Li, W., Guibas, L., and Mitra, N.~J.}
\newblock 2011.
\newblock Exploration of continuous variability in collections of 3d shapes.
\newblock {\em ACM Trans. on Graphics 30}, 4, 33:1--33:10.

\bibitem[\protect\citename{Ovsjanikov et~al\mbox{.} }2012]{Ovsjanikov2012}
{\sc Ovsjanikov, M., Ben-Chen, M., Solomon, J., Butscher, A., and Guibas, L.}
\newblock 2012.
\newblock Functional maps: A flexible representation of maps between shapes.
\newblock {\em ACM Trans. Graph. 31}, 4 (July), 30:1--30:11.

\bibitem[\protect\citename{Parent }1977]{Parent1977}
{\sc Parent, R.~E.}
\newblock 1977.
\newblock A system for sculpting 3-d data.
\newblock {\em SIGGRAPH Comput. Graph. 11}, 2 (July), 138--147.

\bibitem[\protect\citename{Pasko et~al\mbox{.} }1995]{Pasko1995}
{\sc Pasko, A., Adzhiev, V., Sourin, A., and Savchenko, V.}
\newblock 1995.
\newblock Function representation in geometric modeling: concepts,
  implementation and applications.
\newblock {\em The Visual Computer 11}, 8, 429--446.

\bibitem[\protect\citename{Pasko et~al\mbox{.} }2004]{Pasko2004}
{\sc Pasko, G., Pasko, A., and Kunii, T.}
\newblock 2004.
\newblock Space–time blending.
\newblock {\em Computer Animation and Virtual Worlds 15}, 2, 109--121.

\bibitem[\protect\citename{Pauly et~al\mbox{.} }2003]{Pauly2003}
{\sc Pauly, M., Keiser, R., Kobbelt, L.~P., and Gross, M.}
\newblock 2003.
\newblock Shape modeling with point-sampled geometry.
\newblock {\em ACM Trans. Graph. 22}, 3 (July), 641--650.

\bibitem[\protect\citename{Peng et~al\mbox{.} }2004]{Peng2004}
{\sc Peng, J., Kristjansson, D., and Zorin, D.}
\newblock 2004.
\newblock Interactive modeling of topologically complex geometric detail.
\newblock In {\em ACM Transactions on Graphics (TOG)}, vol.~23, ACM, 635--643.

\bibitem[\protect\citename{Poser }2014]{Poser2014}
{\sc Poser}, 2014.
\newblock Smith Micro Software, \texttt{http://my.smithmicro.com}.

\bibitem[\protect\citename{Praun et~al\mbox{.} }2001]{Praun2001}
{\sc Praun, E., Sweldens, W., and Schr\"{o}der, P.}
\newblock 2001.
\newblock Consistent mesh parameterizations.
\newblock In {\em Proceedings of the 28th Annual Conference on Computer
  Graphics and Interactive Techniques}, ACM, New York, NY, USA, SIGGRAPH '01,
  179--184.

\bibitem[\protect\citename{Reeb }1946]{Reeb1946}
{\sc Reeb, G.}
\newblock 1946.
\newblock {Sur les points singuliers d'une forme de Pfaff compl\`{e}tement
  int\'{e}grable ou d'une fonction num\'{e}rique}.
\newblock {\em Comptes Rendus Acad. Sciences 222\/}, 847--849.

\bibitem[\protect\citename{Requicha }1980]{Requicha1980}
{\sc Requicha, A.~G.}
\newblock 1980.
\newblock Representations for rigid solids: Theory, methods, and systems.
\newblock {\em ACM Comput. Surv. 12}, 4 (Dec.), 437--464.

\bibitem[\protect\citename{Sandilands and Komura }2014]{Sandilands2014}
{\sc Sandilands, P., and Komura, T.}
\newblock 2014.
\newblock Model topology change with correspondence using electrostatics.
\newblock In {\em Proceedings of the 20th ACM Symposium on Virtual Reality
  Software and Technology}, ACM, New York, NY, USA, VRST '14, 41--44.

\bibitem[\protect\citename{Savchenko et~al\mbox{.} }1995]{Savchenko1995}
{\sc Savchenko, V.~V., Pasko, A.~A., Okunev, O.~G., and Kunii, T.~L.}
\newblock 1995.
\newblock {Function Representation of Solids Reconstructed from Scattered
  Surface Points and Contours}.
\newblock {\em Computer Graphics Forum\/}.

\bibitem[\protect\citename{Schmidt and Singh }2010]{Schmidt2010}
{\sc Schmidt, R., and Singh, K.}
\newblock 2010.
\newblock Meshmixer: An interface for rapid mesh composition.
\newblock In {\em ACM SIGGRAPH 2010 Talks}, ACM, New York, NY, USA, SIGGRAPH
  '10, 6:1--6:1.

\bibitem[\protect\citename{Schmidt et~al\mbox{.} }2006]{Schmidt2006}
{\sc Schmidt, R., Wyvill, B., Sousa, M.~C., and Jorge, J.~A.}
\newblock 2006.
\newblock Shapeshop: Sketch-based solid modeling with blobtrees.
\newblock In {\em ACM SIGGRAPH 2006 Courses}, ACM, New York, NY, USA, SIGGRAPH
  '06.

\bibitem[\protect\citename{Sebastian et~al\mbox{.} }2001]{Sebastian01}
{\sc Sebastian, T.~B., Klein, P.~N., and Kimia, B.~B.}
\newblock 2001.
\newblock Recognition of shapes by editing their shock graphs.
\newblock In {\em Proc. Int'l Conf. Computer Vision}, 755--762.

\bibitem[\protect\citename{Sederberg and Greenwood }1992]{Sederberg92}
{\sc Sederberg, T.~W., and Greenwood, E.}
\newblock 1992.
\newblock A physically based approach to 2-{D} shape blending.
\newblock In {\em Proc. SIGGRAPH}, 25--34.

\bibitem[\protect\citename{Shapira et~al\mbox{.} }2008]{Shapira2008}
{\sc Shapira, L., Shamir, A., and Cohen-Or, D.}
\newblock 2008.
\newblock Consistent mesh partitioning and skeletonisation using the shape
  diameter function.
\newblock {\em Vis. Comput. 24}, 4 (Mar.), 249--259.

\bibitem[\protect\citename{Sharf et~al\mbox{.} }2006]{Sharf2006}
{\sc Sharf, A., Blumenkrants, M., Shamir, A., and Cohen-Or, D.}
\newblock 2006.
\newblock Snappaste: an interactive technique for easy mesh composition.
\newblock {\em The Visual Computer 22}, 9-11, 835--844.

\bibitem[\protect\citename{Shinagawa et~al\mbox{.} }1991]{Shinagawa1991}
{\sc Shinagawa, Y., Kunii, T., and Kergosien, Y.}
\newblock 1991.
\newblock Surface coding based on morse theory.
\newblock {\em Computer Graphics and Applications, IEEE 11}, 5 (Sept), 66--78.

\bibitem[\protect\citename{Siddiqi et~al\mbox{.} }1999]{Siddiqi1999}
{\sc Siddiqi, K., Shokoufandeh, A., Dickinson, S.~J., and Zucker, S.~W.}
\newblock 1999.
\newblock Shock graphs and shape matching.
\newblock {\em International Journal of Computer Vision 35}, 1, 13--32.

\bibitem[\protect\citename{Sidi et~al\mbox{.} }2011]{Sidi11}
{\sc Sidi, O., van Kaick, O., Kleiman, Y., Zhang, H., and Cohen-Or, D.}
\newblock 2011.
\newblock Unsupervised co-segmentation of a set of shapes via descriptor-space
  spectral clustering.
\newblock {\em ACM Trans. on Graphics 30}, 6, 126:1--126:9.

\bibitem[\protect\citename{Smithe }1990]{Smithe1990}
{\sc Smithe, D.~B.}
\newblock 1990.
\newblock A two-pass mesh warping algorithm for object transformation and image
  interpolation.
\newblock Technical memo \#1030, Industrial Light and Magic.

\bibitem[\protect\citename{Stanculescu et~al\mbox{.} }2011]{Stanculescu2011}
{\sc Stanculescu, L., Chaine, R., and Cani, M.-P.}
\newblock 2011.
\newblock Freestyle: Sculpting meshes with self-adaptive topology.
\newblock {\em Comput. Graph.-UK 35}, 3 (June), 614--622.
\newblock Special Issue: Shape Modeling International (SMI) Conference 2011.

\bibitem[\protect\citename{Surazhsky et~al\mbox{.} }2001]{Surazhsky2001}
{\sc Surazhsky, T., Surazhsky, V., Barequet, G., and Tal, A.}
\newblock 2001.
\newblock Blending polygonal shapes with different topologies.
\newblock {\em Computers \& Graphics 25}, 1, 29 -- 39.
\newblock Shape Blending.

\bibitem[\protect\citename{Szeliski }2010]{Szeliski2010}
{\sc Szeliski, R.}
\newblock 2010.
\newblock {\em Computer Vision: Algorithms and Applications}, 1st~ed.
\newblock Springer-Verlag New York, Inc., New York, NY, USA.

\bibitem[\protect\citename{Takayama et~al\mbox{.} }2011]{Takayama2011}
{\sc Takayama, K., Schmidt, R., Singh, K., Igarashi, T., Boubekeur, T., and
  Sorkine, O.}
\newblock 2011.
\newblock Geobrush: Interactive mesh geometry cloning.
\newblock {\em Computer Graphics Forum 30}, 2, 613--622.

\bibitem[\protect\citename{Tam et~al\mbox{.} }2013]{RegistrationSurvey2013}
{\sc Tam, G. K.~L., Cheng, Z.-Q., Lai, Y.-K., Langbein, F.~C., Liu, Y.,
  Marshall, D., Martin, R.~R., Sun, X.-F., and Rosin, P.~L.}
\newblock 2013.
\newblock Registration of 3d point clouds and meshes: A survey from rigid to
  nonrigid.
\newblock {\em IEEE Transactions on Visualization and Computer Graphics 19}, 7,
  1199--1217.

\bibitem[\protect\citename{Tevs et~al\mbox{.} }2014]{Tevs14}
{\sc Tevs, A., Huang, Q., Wand, M., Seidel, H.-P., and Guibas, L.}
\newblock 2014.
\newblock Relating shapes via geometric symmetries and regularities.
\newblock {\em ACM Trans. on Graphics 33}, 4, 119:1--119:12.

\bibitem[\protect\citename{Turk and O'Brien }1999]{Turk1999}
{\sc Turk, G., and O'Brien, J.~F.}
\newblock 1999.
\newblock Shape transformation using variational implicit functions.
\newblock In {\em Proceedings of the 26th Annual Conference on Computer
  Graphics and Interactive Techniques}, ACM Press/Addison-Wesley Publishing
  Co., New York, NY, USA, SIGGRAPH '99, 335--342.

\bibitem[\protect\citename{van Kaick et~al\mbox{.} }2010]{vanKaick_survey_10}
{\sc van Kaick, O., Zhang, H., Hamarneh, G., and Cohen-Or, D.}
\newblock 2010.
\newblock A survey on shape correspondence.
\newblock {\em Computer Graphics Forum 30}, 6, 1681--1707.

\bibitem[\protect\citename{Wang et~al\mbox{.} }2011]{Wang11}
{\sc Wang, Y., Xu, K., Li, J., Zhang, H., Shamir, A., Liu, L., Cheng, Z., and
  Xiong, Y.}
\newblock 2011.
\newblock Symmetry hierarchy of man-made objects.
\newblock {\em Computer Graphics Forum (Proc. EUROGRAPHICS) 30}, 2, 287--296.

\bibitem[\protect\citename{Weiler }1986]{Weiler86}
{\sc Weiler, K.~J.}
\newblock 1986.
\newblock {\em Topological structures for geometric modeling}.
\newblock PhD thesis, Rensselaer Polytechnic Institute.

\bibitem[\protect\citename{Welch and Witkin }1994]{Welch1994}
{\sc Welch, W., and Witkin, A.}
\newblock 1994.
\newblock Free-form shape design using triangulated surfaces.
\newblock In {\em Proceedings of the 21st Annual Conference on Computer
  Graphics and Interactive Techniques}, ACM, New York, NY, USA, SIGGRAPH '94,
  247--256.

\bibitem[\protect\citename{Weng et~al\mbox{.} }2013]{Weng2013}
{\sc Weng, Y., Chai, M., Xu, W., Tong, Y., and Zhou, K.}
\newblock 2013.
\newblock As-rigid-as-possible distance field metamorphosis.
\newblock {\em Computer Graphics Forum 32}, 7, 381--389.

\bibitem[\protect\citename{Wolberg }1990]{Wolberg1990}
{\sc Wolberg, G.}
\newblock 1990.
\newblock {\em Digital Image Warping}.
\newblock IEEE Computer Society Press, Los Alamitos, CA.

\bibitem[\protect\citename{Wolberg }1998]{Wolberg1998}
{\sc Wolberg, G.}
\newblock 1998.
\newblock Image morphing: a survey.
\newblock {\em The Visual Computer 14}, 8-9, 360--372.

\bibitem[\protect\citename{Wood et~al\mbox{.} }2004]{Wood2004}
{\sc Wood, Z., Hoppe, H., Desbrun, M., and Schr\"{o}der, P.}
\newblock 2004.
\newblock Removing excess topology from isosurfaces.
\newblock {\em ACM Trans. Graph. 23}, 2 (Apr.), 190--208.

\bibitem[\protect\citename{Xie et~al\mbox{.} }2013]{Xie2013}
{\sc Xie, X., Xu, K., Mitra, N.~J., Cohen-Or, D., Gong, W., Su, Q., and Chen,
  B.}
\newblock 2013.
\newblock Sketch-to-design: Context-based part assembly.
\newblock {\em Computer Graphics Forum 32}, 8, 233--245.

\bibitem[\protect\citename{Xu et~al\mbox{.} }2012]{Xu2012}
{\sc Xu, K., Zhang, H., Cohen-Or, D., and Chen, B.}
\newblock 2012.
\newblock Fit and diverse: Set evolution for inspiring 3d shape galleries.
\newblock {\em ACM Trans. Graph. 31}, 4 (July), 57:1--57:10.

\bibitem[\protect\citename{Yin et~al\mbox{.} }2014]{Yin2014}
{\sc Yin, K., Huang, H., Zhang, H., Gong, M., Cohen-Or, D., and Chen, B.}
\newblock 2014.
\newblock Morfit: Interactive surface reconstruction from incomplete point
  clouds with curve-driven topology and geometry control.
\newblock {\em ACM Trans. Graph. 33}, 6 (Nov.), 202:1--202:12.

\bibitem[\protect\citename{ZBrush }2015]{ZBrush2015}
{\sc ZBrush}, 2015.
\newblock Pixologic, \texttt{http://pixologic.com/zbrush}.

\bibitem[\protect\citename{Zhang et~al\mbox{.} }2008]{Zhang08}
{\sc Zhang, H., Sheffer, A., Cohen-Or, D., Zhou, Q., van Kaick, O., and
  Tagliasacchi, A.}
\newblock 2008.
\newblock Deformation-driven shape correspondence.
\newblock {\em Computer Graphics Forum (Proc. SGP) 27}, 5, 1431--1439.

\bibitem[\protect\citename{Zhang et~al\mbox{.} }2010]{Zhang2010}
{\sc Zhang, H., Van~Kaick, O., and Dyer, R.}
\newblock 2010.
\newblock Spectral mesh processing.
\newblock {\em Computer Graphics Forum 29}, 6, 1865--1894.

\bibitem[\protect\citename{Zhao et~al\mbox{.} }2003]{Zhao2003}
{\sc Zhao, Y., Ong, H.-Y., Tan, T.-S., and Xiao, Y.}
\newblock 2003.
\newblock Interactive control of component-based morphing.
\newblock In {\em Proceedings of the 2003 ACM SIGGRAPH/Eurographics Symposium
  on Computer Animation}, Eurographics Association, Aire-la-Ville, Switzerland,
  Switzerland, SCA '03, 339--348.

\bibitem[\protect\citename{Zheng et~al\mbox{.} }2011]{Zheng11}
{\sc Zheng, Y., Fu, H., Cohen-Or, D., Au, O. K.-C., and Tai, C.-L.}
\newblock 2011.
\newblock Component-wise controllers for structure-preserving shape
  manipulation.
\newblock {\em Computer Graphics Forum (Proc. EUROGRAPHICS) 30}, 2, 563--572.

\bibitem[\protect\citename{Zheng et~al\mbox{.} }2013]{Zheng2013}
{\sc Zheng, Y., Cohen-Or, D., and Mitra, N.~J.}
\newblock 2013.
\newblock Smart variations: Functional substructures for part compatibility.
\newblock {\em Computer Graphics Forum 32}, 2pt2, 195--204.

\bibitem[\protect\citename{Zheng et~al\mbox{.} }2014]{Zheng14}
{\sc Zheng, Y., Cohen-Or, D., Averkiou, M., and Mitra, N.~J.}
\newblock 2014.
\newblock Recurring part arrangements in shape collections.
\newblock {\em Computer Graphics Forum (Proc. EUROGRAPHICS) 33}, 2.

\bibitem[\protect\citename{Zhou et~al\mbox{.} }2006]{Zhou2006}
{\sc Zhou, K., Huang, X., Wang, X., Tong, Y., Desbrun, M., Guo, B., and Shum,
  H.-Y.}
\newblock 2006.
\newblock Mesh quilting for geometric texture synthesis.
\newblock {\em ACM Transactions on Graphics (TOG) 25}, 3, 690--697.

\bibitem[\protect\citename{Zhou et~al\mbox{.} }2007]{Zhou2007}
{\sc Zhou, Q.-Y., Ju, T., and Hu, S.-M.}
\newblock 2007.
\newblock Topology repair of solid models using skeletons.
\newblock {\em Visualization and Computer Graphics, IEEE Transactions on 13},
  4, 675--685.

\end{thebibliography}

\end{document}